\documentclass[final,5p,times]{elsarticle}

\usepackage{amssymb}
\usepackage{amsmath}
\usepackage{amsthm}

\usepackage{color}
\usepackage{subcaption}
\usepackage{cleveref}
\biboptions{sort&compress}

\newcommand{\vmes}{{\bf v}}

\newcommand{\kbt}{k_BT}
\newcommand{\fij}{F_{ij}}

\newcommand{\rbij}{{\bf r}_{ij}}
\newcommand{\rbi}{{\bf r}_{i}}
\newcommand{\rbj}{{\bf r}_{j}}

\newcommand{\Aij}{A_{ij}}
\newcommand{\Bij}{B_{ij}}

\newcommand{\eij}{{{\bf e}_{ij}}}

\newcommand{\wbbarij}{{\bf \overline{W}}_{ij}}
\newcommand{\wbij}{{\bf {W}}_{ij}}
\newcommand{\fa}{\mathcal{A}}
\newcommand{\fb}{\mathcal{B}}

\newcommand{\der}{\text{d}}

\definecolor{green}{RGB}{1,129,30}

\newcommand{\ratfo}{k_f}
\newcommand{\ratde}{k_d}

\newcommand{\tact}{\lambda_{\text{thix}}}
\newcommand{\tsnd}{\tau_{\text{c}}}
\newcommand{\tconv}{\tau_{\text{flow}}}
\newcommand{\tsig}{\tau_{\sigma}}
\newcommand{\tth}{\tau_{\text{t}}}
\newcommand{\tvis}{\tau_{\text{vis}}}

\journal{Computer Physics Communications}

\begin{document}

\begin{frontmatter}



\title{Computational Modelling of Thixotropic Multiphase Fluids}


\author[label1,label2]{Andres Santiago Espinosa-Moreno\corref{cor2}} 
\cortext[cor2]{aespinosa@bcamath.org}

\author[label1]{Nicol\'as Moreno\corref{cor1}} 
\cortext[cor1]{nmoreno@bcamath.org}

\author[label1,label3,label4]{Marco Ellero} 

\affiliation[label1]{organization={Basque Center for Applied Mathematics (BCAM)},
            addressline={Alameda de Mazarredo, 14}, 
            city={Bilbao},
            postcode={48400}, 
            state={Basque Country},
            country={Spain}}

\affiliation[label2]{organization={University of the Basque Country/Euskal Herriko Unibertsitatea},
            addressline={Barrio Sarriena}, 
            city={Leioa},
            postcode={48940}, 
            state={Basque Country},
            country={Spain}}
            
\affiliation[label3]{organization={Basque Foundation for Science (IKERBASQUE)},
            addressline={Calle de Maria Diaz de Haro 3}, 
            city={Bilbao},
            postcode={480013}, 
            state={Basque Country},
            country={Spain}}
            
\affiliation[label4]{organization={Complex Fluids Research Group, Department of Chemical Engineering, Faculty of Science and Engineering, Swansea University},
addressline={ SA1 8EN}, 
            city={Swansea},
            country={UK}}

\begin{abstract}
Multiphase systems are ubiquitous in engineering, biology, and materials science, where understanding their complex interactions and rheological behavior is crucial for advancing applications ranging from emulsion stability to cellular phase separation. This study presents a numerical methodology for modeling thixotropic multiphase fluids, emphasizing the transient behavior of viscosity and the intricate interactions between phases. The model incorporates phase-dependent viscosities, interfacial tension effects, and the dynamics of phase separation, coalescence, and break-up, making it suitable for simulating systems with complex flow regimes. A key feature of the methodology is its ability to capture thixotropic behavior, where viscosity evolves over time due to microstructural changes induced by shear history. This approach enables the simulation of aging and recovery processes in materials such as gels, emulsions, and biological tissues. The model is rigorously validated against benchmark cases, demonstrating its accuracy in predicting multiphase systems under static and dynamic conditions. Subsequently, the methodology is applied to investigate systems with varying levels of microstructural evolution, revealing the impact of thixotropic dynamics on overall system behavior. The results provide new insights into the time-dependent rheology of multiphase fluids and highlight the versatility of the model for applications in industrial and biological systems involving complex fluid interactions.
\end{abstract}




\end{frontmatter}



\section{Introduction}\label{sec1}

Multiphase systems are crucial to engineering, biology, and materials science, where the existence of complex microstructural dynamics often lead to non-trivial rheological behavior, such as shear-thinning or thixotropic properties. In industrial applications, understanding emulsion and foam stability is critical for improving product quality and efficiency \cite{dickinson2020advances}. In biology, liquid-liquid phase separation within cells organizes cellular components and influences diseases like neurodegenerative disorders \cite{shin2017liquid, alberti2021biomolecular, spannl2019biomolecular, hyman2014liquid}. Similarly, interactions among blood components, such as plasma and red blood cells, are essential for processes like clotting under varying physiological conditions \cite{fogelson2015fluid, smith2015all, ataullakhanov1998spatiotemporal}. In materials science, understanding phase interactions is crucial for designing smart materials, composites, and alloys with desired properties \cite{huang2021interphase, pramod2015aluminum}. Accurately modeling these systems, including phase separation dynamics and time-dependent material properties \cite{rossi2022sph}, remains challenging, driving the development of advanced grid-based and mesh-free methods.

Grid-based methods such as the Finite Element \cite{soloveichik2022method,molina2023finite,mavridis1987finite}, Lattice Boltzmann Method (LBM) \cite{yan2011lbm,petersen2021lattice,reis2022lattice},  methods based on the Volume of Fluid (VOF) \cite{dolai2025hybrid,garoosi2022numerical,wang2022volume}, Level Set (LS) \cite{jettestuen2021locally, doherty2023stabilised, lyras2023conservative} or Immersed Boundary Method (IBM) \cite{patel2017coupled,zhenga2017coupled}, are widely used for tracking interfaces and capturing phase interactions. These methods are particularly effective for problems involving sharp interfaces. However, they can struggle with dynamic interface tracking, requiring significant computational resources for high-resolution simulations. Additionally, spurious non-physical velocity fields near the interfaces (resulting from discretization errors) can change the shape or stability of the simulated phases \cite{leclaire2015approach, xu2017lattice, leclaire2016modeling}. Mesh-free methods, such as Smoothed Particle Hydrodynamics (SPH) \cite{Ellero2007,tartakovsky2016pairwise,hu2009constant,le2025smoothed,liu2010smoothed} and Dissipative Particle Dynamics (DPD) \cite{GrootAndWarren1997, espanol2017perspective, gavrilov2011microphase,tiwari2006dissipative}, have gained popularity for their ability to handle complex geometries and dynamic interfaces without the need for explicitly tracking the interfaces. These methods are particularly advantageous for simulating multiphase flows with large deformations\cite{liu2010smoothed} and complex interactions\cite{moreno2020}. However, they can suffer from numerical instabilities, especially at interfaces, where large differences in viscosity and density can generate pressure oscillations and unphysical artifacts, such as gaps or particle mixing at the interface \cite{monaghan2013simple, chen2015sph, meng2020multiphase,yang2022consistent, guo2024smoothed}. Additionally, implementing accurate boundary conditions at walls and interfaces for fluids with varying viscosities is complex and can result in errors affecting flow behavior near these regions.

Smoothed dissipative particle dynamics (SDPD) \cite{espanol2003smoothed,ellero2018everything} is a mesoscopic technique that combines the discretization scheme of SPH and the consistent thermal fluctuations of DPD, effectively reproducing the fluctuating Navier-Stokes equations. SDPD has been successfully applied to model a range of synthetic~\cite{vazquez2012sph,simavilla2022mesoscopic,bian2012multiscale,moreno2023,echeverria2025} and biological systems~\cite{moreno2013multiscale, zohravi2023computational, zohravi2025mesoscale, moreno-chaparro2025}. Recently, Lei et al.~\cite{lei2016smoothed} developed a SDPD scheme suitable to model multiphase systems incorporating surface tension effects via a pairwise interaction forces.  This scheme was able to capture complex phenomena such as droplet coalescence and interface stabilization in the presence of thermal noise, showing results consistent with theoretical predictions and experimental data.  The method's ability to incorporate thermal fluctuations makes it particularly suitable for studying micro to nanoscale systems where thermal fluctuations can  play a role in phase interactions and dynamics. However, the application of SDPD to multiphase flows with complex rheological behavior, such as thixotropic materials, which is crucial for accurately simulating complex fluids like gels or emulsions that exhibit microstructural changes during flow, remains an emerging field of research.  

In this work, we introduce a comprehensive SDPD model designed to address the challenges inherent in modeling multiphase systems with complex rheological behavior. The proposed model incorporates several key features to accurately capture the dynamics of such systems. First, it incorporates interfacial tension effects, which are essential for understanding interactions at phase boundaries, such as those observed in emulsions. Second, the model accounts for the dynamics of phase separation, including phenomena like coalescence and break-up, which are critical for predicting the stability and structural evolution of multiphase systems under varying flow conditions. Third, it explicitly accounts for phase-dependent viscosities, enabling precise representation of systems where fluids or materials exhibit distinct flow properties. Additionally, the methodology integrates a thixotropic viscosity model to capture time-dependent rheological behavior, allowing for the simulation of aging and recovery processes in materials such as gels and emulsions. By incorporating microstructural dynamics, the model provides a versatile framework for studying systems that undergo structural changes during flow.

The present manuscript is organized as follows. In the first section, we introduce our methodology and numerical model. In the second section, we validate the robustness and accuracy of our implementation for both Newtonian and Non-Newtonian (thixotropic) fluids in static and dynamic states. We further illustrate the flexibility of the methodology for various applications in biology and microfluidics, including liquid-liquid phase separation, thixotropic emulsions flow, and complex-microfluidic geometries to control merging and splitting in multiphase flows. Finally, we provide the main conclusions and recommendations arising from this work.

\section{Numerical methodology}\label{sec:met}

In SDPD, a fluid is discretized using particles with a volume $V_i$, leading to a particle number density $d_i = 1/V_i = \sum_j W(r_{i j},h)$. $W(r_{i j},h)$ is a kernel function that depends on the distance $r_{i j} = |{\bf{r}}_i - {\bf{r}}_j|$ between particles $i$ and $j$, has a finite support $h$ and is normalized to one.  The evolution of the particle position is given by the equation ${\der {\bf{r}}_i}/{\der t} = \vmes_i$, where $\vmes_i$ is the velocity of the $i$-th particle. The momentum's stochastic differential equation is given by
\begin{align}
m_i\frac{\der \vmes_i}{\der t}  &= \sum_j \left[ \frac{P_i}{d_i^2} + \frac{P_j}{d_j^2}\right] \fij \rbij \\
&-\sum_j \left[\fa_{ij} \vmes_{ij} +\fb_{ij} (\vmes_{ij}\cdot\eij)\eij \right] \frac{\fij}{d_id_j} \nonumber \\
&+\sum_j \left(\Aij \der \wbbarij + \Bij \frac{1}{D}\text{tr}[ \der \wbij] \right) \cdot \frac{\eij}{\der t},
\label{eq:deterministic}
\end{align}
where $P_i$ and $P_j$ are the pressure of particles $i$ and $j$, respectively. The pressure of each particle is determined by a suitable density-dependent equation of state $P = f(\rho)$, as discussed later. The term $m_i = \rho {\Delta x}^D$, denotes the mass of the particle that depends on the density $\rho$, the equilibrium interparticle distance $\Delta x$, and the dimension of the system $D$. The term $F_{i j} =- \nabla W(r_{i j},h)/r_{i j}$ is a positive function introduced to account for the interaction forces between particles. Here, we adopt a Lucy's Quartic kernel~\cite{lucy1977numerical} of the form $W(r) = \mathrm{w_0} (1+3{r}/{h}) (1-{r}/{h})^3$ if $r/h \leq 1$ and $W(r) = 0$ if $r/h > 1$, where $\mathrm{w_0}  = 5/(\pi h^2)$ for two dimensions and $\mathrm{w_0}  = 105/(16 \pi h^3)$ for three dimensions. The second term in \eqref{eq:deterministic} corresponds to viscous contributions, where $\vmes_{ij} = \vmes_i - \vmes_j$ is the relative velocity between particles, $\eij = \rbij/|\rbij|$ is the unit vector, $\fa_{ij}$ and $\fb_{ij}$ are friction coefficients related to the pairs shear viscosity ($\eta_{ij}$) and bulk viscosity ($\zeta_{ij} = \eta_{ij} (2D-1)/D$), with $\fa_{ij}={(D+2)\eta_{ij}}/{D}-\zeta_{ij}$ and $\fb_{ij} = (D+2)(\zeta_{ij}+{\eta_{ij}}/{D})$. Here, we consider the viscosity between particles $i$ and $j$ expressed as $\eta_{ij} = 2 (\eta_i \eta_j)/(\eta_i + \eta_j)$. The last term in \eqref{eq:deterministic} incorporates thermal fluctuations into the momentum balance by including a matrix of independent increments of the Wiener process, $d\wbij$ and its traceless symmetric part $d\wbbarij$ given by
\begin{equation}
    d \wbbarij = \frac{1}{2} [d \wbij + d \wbij^T ] - \frac{\delta^{\alpha \beta}}{D} \text{tr}[d \wbij].
\end{equation}
The parameters $A_{ij}$ and $B_{ij}$ are the amplitude of the noise, related to the friction coefficients ($\fa_{ij}$ and $\fb_{ij}$) and, in order to satisfy the fluctuation-dissipation theorem, they are given by
\begin{align}
    A_{ij} &= \left[4 \ k_B T \ \fa_{ij}  \ \frac{F_{i j}}{d_i d_j} \right]^{-1/2} \\
    B_{ij} &= \left[4 \ k_B T \ \left(\fb_{ij} - \fa_{ij} \frac{D-2}{D}\right)  \ \frac{F_{i j}}{d_i d_j} \right]^{-1/2},
\end{align}
where $k_B$ is the Boltzmann constant and $T$ temperature.

\subsection{Thixotropic model}

When exposed to continuous shear stress, some non-Newtonian fluids can exhibit a time-dependent shear-thinning or thixotropic behavior, which is characterized by a gradual decrease in viscosity followed by a recovery upon removal of the stress. To capture this behavior, we can follow the phenomenological description given by Le-Cao et al. 
\cite{le2020microstructure},  where viscosity of a fluid  is given by 
\begin{equation}
\eta(t,x) = \eta_{\infty} \left[ 1 + \alpha f(t,\dot{\gamma}(x)) \right] 
\label{eq:etavar}
\end{equation}
where $\eta(t)$ is the time-dependent shear viscosity, $\eta_\infty$ is the limiting lowest viscosity, $\alpha$ is a constitutive constant, and $f \in [0 : 1]$ is a microstructural scalar parameter describing the current state of the microstructure, continuously from fully destroyed ($f=0$) to completely developed ($f=1$) \cite{rossi2022sph}. Notice from \Cref{eq:etavar}, that for a fully developed microstructure, $\alpha$ indicates the maximum viscosity $\eta_{\text{max}} = \eta_{\infty}(1+\alpha)$. The time evolution of $f$ is given by 

\begin{equation}
\dot{f} = \ratfo - (\ratfo + \ratde \dot{\gamma})f    
\label{eq:fdot}
\end{equation}

where $\ratfo$ is a constant that represents the rate of formation of the microstructure, and $\ratde \dot{\gamma}$ the rate of destruction of the microstructure under shear conditions. The strain-rate tensor $\dot{\boldsymbol{\gamma}}$ is given by $\dot{\boldsymbol{\gamma}} = (\nabla \vmes ) + (\nabla \vmes^T)$. The
strain-rate tensor second invariant $\dot{\gamma}$ appearing in \Cref{eq:etavar} can be evaluated as:

\begin{equation}
    \dot{\gamma} = \sqrt{\frac{1}{2} \it \Pi_{\dot{\gamma}}} = \left[ \frac{1}{2}  \{ \dot{\boldsymbol{\gamma}} : \dot{\boldsymbol{\gamma}} \}  \right]^{1/2}
\end{equation}

The velocity gradient tensor $\nabla \vmes$ is calculated in SDPD as
\begin{equation}
    (\nabla \vmes)^{\alpha' \beta' }_i = \sum_j \frac{(\vmes^{\alpha'}_i - \vmes^{\alpha'}_j)(\rbi^{\beta'} - \rbj^{\beta'})}{d_j r_{ij}} \frac{\partial W(r_{ij},h)}{\partial r} ,
\end{equation}
where $\alpha',\beta'$ are the column and row indices in the matrix tensor and $\partial W(r_{ij},h)/\partial r$ is the gradient of the kernel. By considering the ratio between the formation and destruction and the characteristic time for microstructure formation, we can introduce the parameter $\beta = \ratde/\ratfo$ and the thixotropic time scale $\lambda_{\text{thix}} = 1/\ratfo$. Accordingly, the temporal evolution can be rewritten as $\dot{f} = [1 - (1 + \beta \dot{\gamma})f] \lambda_{\text{thix}}^{-1}$ and, after time integration, the scalar parameter $f$ can be calculated as \cite{rossi2022sph}
\begin{equation}\label{thixo_1}
f(t) = \frac{1}{1 + \beta\dot{\gamma}} \left( 1 - \exp^{-{(1 + \beta\dot{\gamma})t}/{\lambda_{\text{thix}}}} \right) + f_0 \exp^{-{(1 + \beta\dot{\gamma})t}/{\lambda_{\text{thix}}}},
\end{equation}
where $f_0$ is the initial value of the scalar parameter $f$. Similar to the viscosity between a pair of particles $i$ and $j$, any pair parameter $X_{ij}$ of the thixotropic model is approximated by $X_{ij} = 2 (X_i X_j)/(X_i + X_j)$. For instance, we have for the limiting viscosity $(\eta_\infty)_{ij} = 2 ((\eta_\infty)_i (\eta_\infty)_j)/((\eta_\infty)_i + (\eta_\infty)_j)$, and similarly for the rest of factors. A similar microstructural SPH approach has been recently considered to model also discontinuous shear thickening in flow of dense suspensions \cite{angerman2024microstructural,angerman2025numerical}.

\subsection{Multiphase Model}

Interfacial tension can be included in the model using an additional pairwise interaction force $\mathbf{F}^{int}_{ij}$, to the momentum balance \eqref{eq:deterministic}. This interaction force takes the form \cite{tartakovsky2016pairwise} 
\begin{equation}
\mathbf{F}^{int}_{ij} = \mathbf{F}^{int}(\mathbf{r}_{ij}) = - \phi (r_{ij}) s_{ij}  \eij,
\label{eq:pairwise}
\end{equation}
where $\phi$ and $s_{ij}$ determine the functional form of the potential and the strength, respectively. This force is short-range repulsive and long-range attractive \cite{lei2016smoothed}. The shape function, $\phi(r_{ij})$, is given by
\begin{align}
    \phi(r_{ij}) &= r_{ij} \left[ -A e^{- {r^2_{ij}}/{2 \epsilon_0^2}} + e^{- {r^2_{ij}}/{2 \epsilon^2}} \right],
\end{align}
where $A$ represents the magnitude of the ratio between repulsive and attractive forces, and $\epsilon$ and $\epsilon_0$ are compact-support dependent functions. For two different phases 1 and 2 the factor $s_{ij}$ is related to the domains $\Omega_1$ and $\Omega_2$ as
 \begin{equation}
s_{ij} = 
\begin{cases} 
s_{11} & \text{if } \mathbf{r}_i \in \Omega_1 \text{ and } \mathbf{r}_j \in \Omega_1, \\
s_{22} & \text{if } \mathbf{r}_i \in \Omega_2 \text{ and } \mathbf{r}_j \in \Omega_2,
\\
s_{12} & \text{if } \mathbf{r}_i \in \Omega_1 \text{ and } \mathbf{r}_j \in \Omega_2, 
\end{cases}
\label{eq:sij}
\end{equation}
where 
\begin{equation}
s_{11} = s_{22} = 10^k s_{12} = \frac{1}{1-10^{-k}} d^{-2}_{eq} \frac{\sigma_0}{\sqrt{2 \pi} (-A \epsilon_0^5 + \epsilon^5)} 
\label{eq:sij2}
\end{equation}
Here, $k$ is a constant that must be greater than 1, $\sigma_0$ is a macroscopic surface tension and $d_{eq} = 1/\Delta x^{D}$ is the equilibrium particle number density.

Following Tartakovsky and Panchenko, we use $\epsilon= h/3.5$, $\epsilon_0=\epsilon/2$ and $A = (\epsilon/\epsilon_0)^3$. This choice ensures that the virial pressure ($P_v = - \pi (-A \epsilon_0^4 + \epsilon^4) d_{eq}^2 s_{11}$) is negative, which is necessary for the stability of the interfaces~\cite{tartakovsky2016pairwise}.  However, we must note, that the presence of a negative virial pressure can also lead to numerical instabilities in the system, especially when the system is not in equilibrium. In order to prevent this viral pressure affects the global system pressure, we introduce an equation of state fo the form
\begin{equation}
P = \frac{c_s^2 \rho_0}{7} \left[ \left( \frac{\rho}{\rho_0} \right) ^7 -1 \right]  + P_b, 
\label{eq:pressure}
\end{equation}
where $\rho_0$ is the equilibrium density for the system, $c_s$ is the speed of sound, and $P_b = -P_v = - \pi (-A \epsilon_0^4 + \epsilon^4) d_{eq}^2 s_{11}$ is a background pressure that ensures $P>0$. This component balance the system and avoid spurious effects of global negative pressure. It is important to note that the presence of this background pressure can also influence the dynamics of the system, especially in flow conditions. In the Results section, we show that the choice of $P_b$ does not introduce artifacts into the simulations, under different dynamic conditions. 

In our simulations, we adopt the well-known Velvet-velocity algorithm \cite{allen1987computer} for the temporal evolution of the position and velocity of the particles. We use a time step $\Delta t$ that ensure numerical stability\cite{bian2012multiscale}, satisfying both: $\Delta t \leq 0.125 h^2\rho/\eta$ (the viscous diffusion criteria \cite{morris1997modeling}) and $\Delta t\leq 0.25 h/c_s$ (the Courant-Friedrichs-Lewy (CFL) condition \cite{courant1928}).

\subsection{Relevant time scale}

In multiphase thixotropic systems their behavior can be influenced by the interplay of various temporal scales. To streamline the analysis and interpretation of our numerical results, we introduce a set of relevant time scales.  The sonic time scale $\tsnd = R/c_s$ represents the time it takes for sound waves to propagate across a characteristic length scale $R$. The viscous dissipation time scale $\tvis = \rho R^2/\eta$, which characterizes the time required for momentum to diffuse across a characteristic length scale $R$ (e.g., droplet radius). The characteristic flow time scale $\tconv = R/U$ represents the time it takes for a fluid element to traverse a characteristic length scale $R$ at a velocity $U$. A time scale associated with interfacial tension effects is given by $\tsig = \eta R/\sigma$, which characterizes the time it takes for surface tension to influence the flow over a characteristic length scale $R$. The thermal fluctuations time scale can be estimated from the thermal velocity $v_{th} = \sqrt{k_B T/m}$ as $\tth = R/v_{th}$, which represents the time it takes for thermal fluctuations to affect the motion of particles over a characteristic length scale $R$. The following hierarchy of time scales holds for realistic multiphase thixotropic systems $\tsnd \ll \tvis, \tsig \ll \tconv, \tact, \tth$. This hierarchy indicates that sound waves propagate faster than viscous diffusion and interfacial tension, which in turn is faster than convective transport, thermal fluctuations, and microstructural evolution. A summary of the relevant time scales is compiled in Table \ref{tab:timescales}. Our goal is to investigate the interplay between $\tsig$, $\tconv$, $\tact$, and $\tth$ time scales in various multiphase thixotropic systems.

From the ratio between these time scales, we can define several dimensionless numbers that characterize the behavior of multiphase thixotropic systems. The Reynolds number is given by $\text{Re} = \tvis/\tconv = \rho U R/\eta$, which represents the ratio of inertial to viscous forces. The Capillary number is defined as $\text{Ca} = \tconv/\tsig = \eta U/\sigma$, which characterizes the relative importance of viscous forces to surface tension forces. The Peclet number is given by $\text{Pe} = \tth/\tconv = UR/D_{th}$, where $D_{th}$ is the thermal diffusivity, representing the ratio of convective to diffusive transport. Following \cite{mujumdar2002transient,larson2019review,rossi2022sph} we also defined the so called Thixotropic number as $\text{Th} = (\ratde /\ratfo)(1/\tconv)$, which evaluates the interplay between the destruction (due to flow) and formation of the microstructure. Where, $\text{Th}>>1$ indicates that microstructural destruction due to the flow is faster than formation, while $\text{Th}<<1$ indicates the opposite. In this work, we focus on conditions where the times scales of formation and destruction of the microstructure overlap $\text{Th}\simeq1$.
\begin{table*}[t]
\centering
\begin{tabular}{|c|c|c|}
\hline
\textbf{Time Scale} & \textbf{Definition} & \textbf{Dimensional Analysis} \\
\hline
Viscous Diffusion Time Scale & $\displaystyle \tvis = \frac{\rho R^2}{\eta}$ 
& $\frac{[M L^{-3}] [L^2]}{[M L^{-1} T^{-1}]} = [T]$ \\
\hline
Sonic Time Scale & $\displaystyle \tsnd = \frac{R}{c_s}$ 
& $\frac{[L]}{[L T^{-1}]} = [T]$ \\
\hline
Convective Time Scale & $\displaystyle \tconv = \frac{R}{U}$ 
& $\frac{[L]}{[L T^{-1}]} = [T]$ \\
\hline
Interfacial Tension Time Scale & $\displaystyle \tsig = \frac{\eta R}{\sigma}$ 
& $\frac{[M L^{-1} T^{-1}] [L]}{[M T^{-2}]} = [T]$ \\
\hline
Thermal Fluctuation Time Scale & $\displaystyle \tth = \frac{R}{v_{th}}$ 
& $\frac{[L]}{[L T^{-1}]} = [T]$ \\
\hline
Thixotropic Time Scale & $\displaystyle \tact = \frac{1}{\ratfo}$ 
& $[\ratfo] = [T^{-1}] \;\Rightarrow\; \tact = [T]$ \\
\hline
\end{tabular}
\caption{Relevant time scales in multiphase thixotropic systems with their dimensional analysis, confirming that each has dimensions of time [T].}
\label{tab:timescales}
\end{table*}

\section{Results and Discussion}

\subsection{Validation of the numerical methodology}
\begin{figure*}[t!]
\centering
     \begin{subfigure}[b]{0.32\textwidth}
         \centering
     \includegraphics[width=\textwidth]{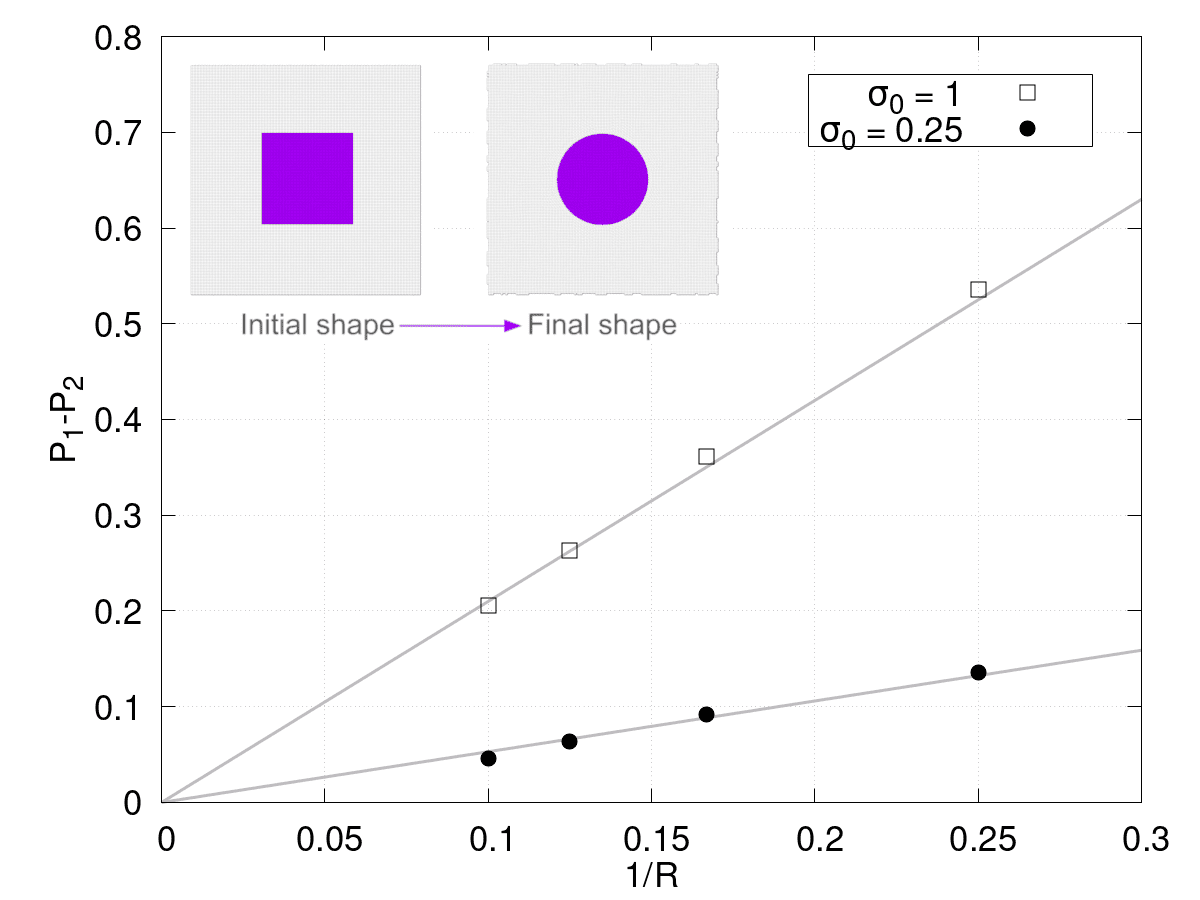}
         \caption{Pressure difference versus curvature $1/R$}
         \label{fig:1-a}
     \end{subfigure}
     \begin{subfigure}[b]{0.32\textwidth}
         \centering
     \includegraphics[width=\textwidth]{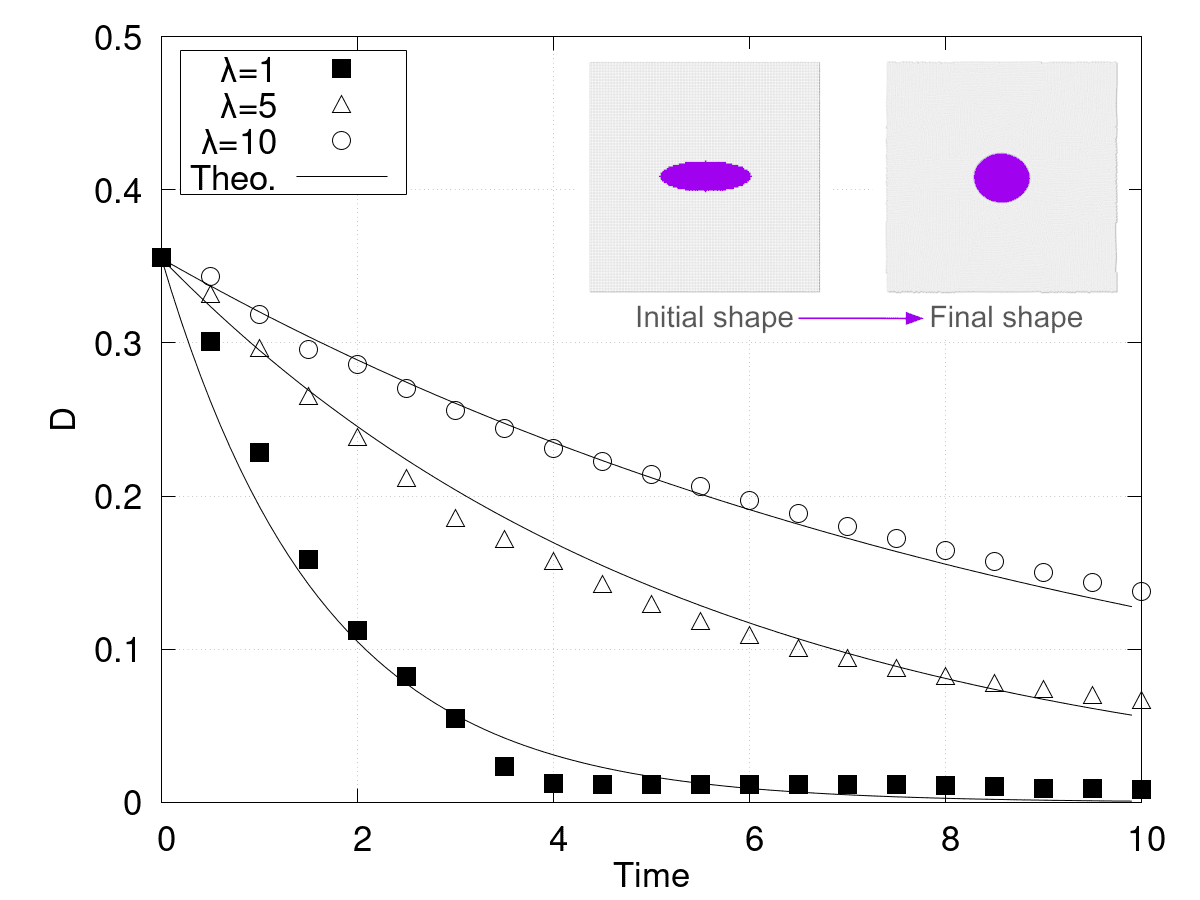}
         \caption{Retraction of a stretched droplet}
         \label{fig:1-b}
     \end{subfigure}
     \begin{subfigure}[b]{0.32\textwidth}
         \centering
     \includegraphics[width=\textwidth]{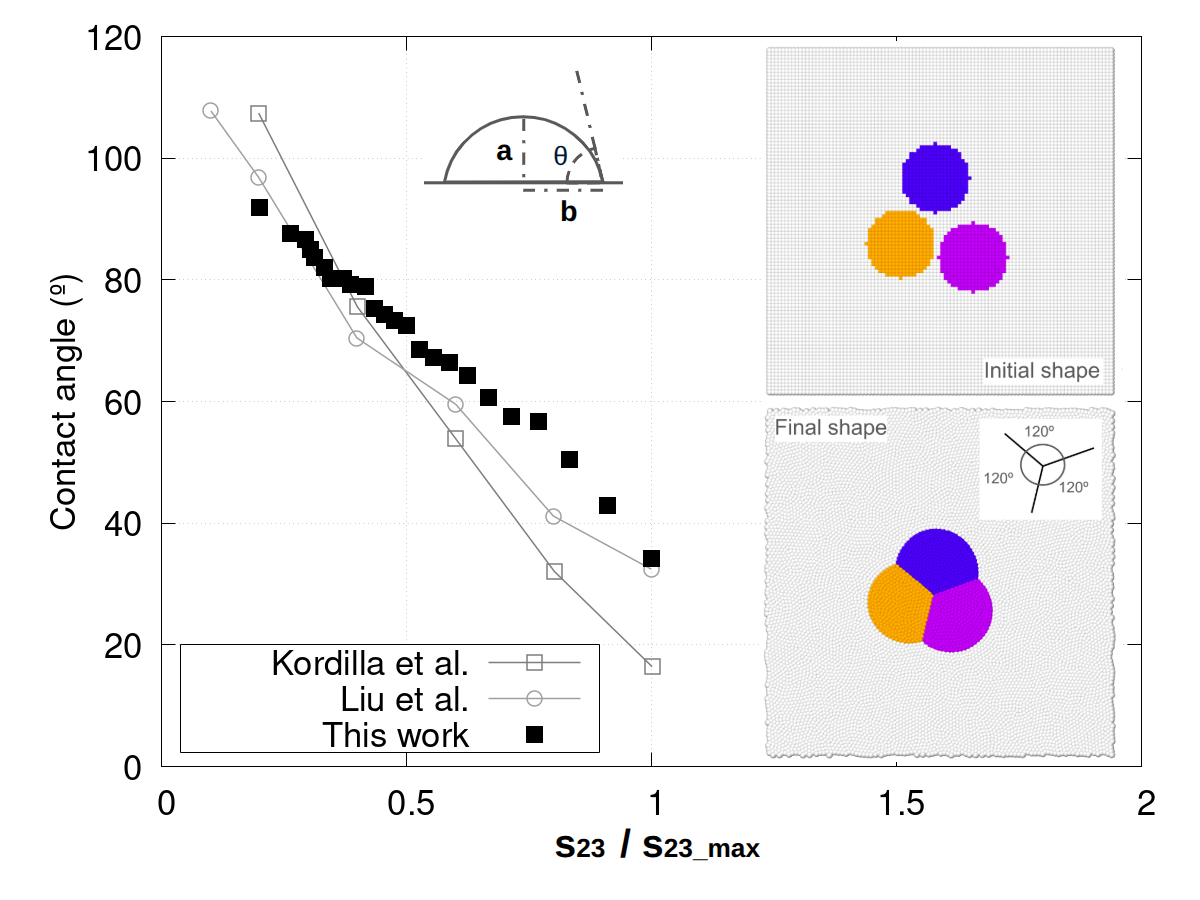}
         \caption{Static contact angle}
         \label{fig:1-c}
     \end{subfigure}     
\caption{Validation of the methodology for multiphase static cases of (a) droplet surface tension, (b) retraction of a stretched droplet and (c) static contact angle between different phases.}
\label{fig:1}
\end{figure*}

We conduct a series of benchmark simulations to ensure the accuracy and robustness of our implementation. First addressing the correct description of the surface tension and contact angles in static and dynamic cases. Then, we validate the thixotropic model by simulating a transient viscosity case. The results of these simulations are compared with theoretical predictions and previous numerical results. Unless otherwise indicated, all the simulations presented use an initial interparticle distance $\Delta x=0.2$. 

\subsubsection{Static validation: Surface tension and contact angles} 

As is customary \cite{tartakovsky2005modeling,tartakovsky2006pore,tartakovsky2009lagrangian,kordilla2013smoothed,tartakovsky2016pairwise,lei2016smoothed}, we initially validate the consistency of our numerical implementation to reproduce the surface tension ($\sigma$) of a spherical droplet in a fluid  \textit{via} the Young Laplace equation, $ \sigma = R(P_d - P_f)/2$, where $R$ is the droplet radius, and $P_d$ and $P_f$ are the pressure of the droplet and external fluid, respectively. For these simulations, we start with a square-shaped particle array of different sizes, that evolve in time to circular droplets of varying size. The square simulation box has a length $L/\Delta x=150$. In \Cref{eq:sij2}, we use an input $\sigma_0 = 0.25$ and $\sigma_0 = 1$, with an equilibrium particle density $n_{eq} = 25$, viscosity $\eta=1$ for both droplet and continuous phase and $k_BT = 0$. These parameters lead to an interfacial tension time scale $\tsig = \{ 1,4\}$ for each of the $\sigma_0$ evaluated, considering a normalized radius droplet with unit length as the characteristic length, i.e., $R=1$. In \Cref{fig:1-a}, we present the evolution of the pressure difference as a function of curvature ($1/R$). Consistent with the Young-Laplace equation, the slopes obtained from the linear fitting, $0.53$ and $2.08$ corresponds to $\sim 2\sigma$. For these surface tension output values, it can be calculated that $\tsig = \{ 0.96,3.78\}$ showing the consistency of the proposed methodology for reproducing this time scale.

We further validate the consistency of our method, measuring the deformation of an initially elliptical droplet as  it retracts to a circular shape due to interfacial tension effects \cite{luciani1997interfacial,guido1999measurement,mo2000new,pan2015studies}. The initial ellipse (at $t=0$) is defined by the semi-major axis $I_A$ and the semi-minor axis $I_B$, defining the so called Taylor deformation parameter \cite{taylor1934formation} ${\mathcal{D}}_0 = (I_A-I_B)/(I_A+I_B)$. The retraction process is governed by the balance between surface tension and viscous forces. The evolution of the of the Taylor parameter $\mathcal{D}$ can be described by \cite{luciani1997interfacial,guido1999measurement,mo2000new} 
\begin{equation}\label{eq:taylor}
    \mathcal{D} = \mathcal{D}_{0} \exp\left( -\frac{\sigma}{\eta R} \frac{40 (\lambda + 1)}{(2\lambda + 3)(19\lambda + 16)} t \right),
\end{equation}
where $R$ is the final circle radius, $\lambda=\eta_d/\eta$ the ratio between the viscosity of the disperse phase $\eta_d$ and the continuous phase $\eta$, and $t$ the stretching time. The system is confined in a square simulation box of side length $L/\Delta x=50$. Here, we use three different viscosity ratios $\lambda = [1,5,10]$, and use Principal Component Analysis (PCA) to compute the eigenvalues and eigenvectors of the droplets particles, providing the magnitude and orientation of the ellipse's principal axes. In \Cref{fig:1-b}, we compare our results for the time evolution of $\mathcal{D}$ with \eqref{eq:taylor}, evidencing a good agreement between the numerical and theoretical results. 

We now validate the model's ability to consistently capture contact angle variations. In \eqref{eq:pairwise}, the strength factor $s_{xx}$ (with $xx$ denoting the phase pair, e.g., $1$, $2$, ..., $n$), governs interfacial interactions and equilibrium configurations \cite{tartakovsky2016pairwise}. To illustrate, we consider a system of four phases where phase $1$ contains droplets of phases $2$, $3$, and $4$, initially non-contacting, as shown in detail in  \Cref{fig:1-c}. We consider a square box for the simulation, where the length of each side is $L/\Delta x=100$. Interaction strengths are defined as $s_{11} = s_{22} = s_{33} = s_{44}$ and $s_{12} = s_{13} = s_{14} = s_{23} = s_{24} = s_{34}$. Under these conditions, the system evolves into an equilibrium configuration with a symmetric triple contact angle of $120^\circ$ between each pair of interfaces. Similarly, varying the interaction strengths  should enable us to control the static contact angles on solid surfaces\cite{kordilla2013smoothed,li2018smoothed}. In \Cref{fig:1-c}, we present the contact angle variation between a fluid (1), a droplet (2) and a solid surface (3). To estimate the contact angle, we measure the droplet height $a_d$ and width $2b_d$, such that $\theta = {\pi}/{2} - \arcsin({b_d^2-a_d^2}/{b_d^2+a_d^2})$~\cite{tartakovsky2016pairwise}.  To obtain angles between $0^o$ and $90^o$ it is necessary that the interaction force between fluid-droplet be less than the interaction force between droplet-solid ($s_{13}<s_{23}$). In the other hand, for $90^o$ to $180^o$, it is necessary that the interaction force between fluid-droplet be greater than the interaction force between droplet-solid ($s_{13}>s_{23}$). We compare the measured angles between 0 and 90 plotted against the droplet-solid interaction force ($s_{23}$) normalized to its maximum value ($s_{23,max}$). In \Cref{fig:1-c}, we show that the measured variation in the contact angle is consistenty with previously reported numerical results ~\cite{kordilla2013smoothed,li2018smoothed}. This static validation confirms that the implemented multiphase scheme reliably captures interfacial behavior and surface tension effects across fluid--fluid and fluid--solid boundaries.

\subsection{Dynamic validation: Poiseuille and shear flow} 
\begin{figure*}[t!]
\centering
     \begin{subfigure}[b]{0.34\textwidth}
         \centering
     \includegraphics[width=\textwidth]{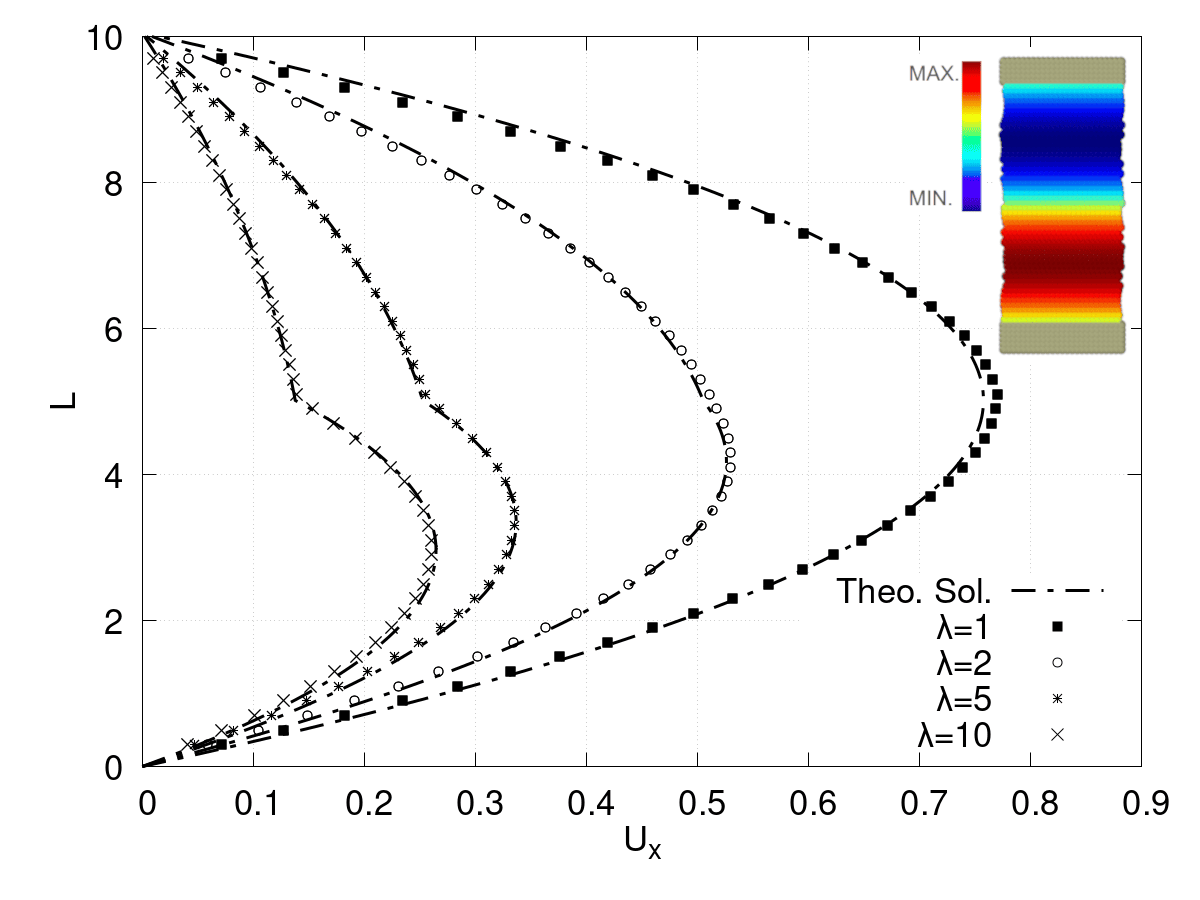}
         \caption{Multiphase flow at different $\lambda$}
         \label{fig:2-a}
     \end{subfigure}
     \begin{subfigure}[b]{0.34\textwidth}
         \centering
     \includegraphics[width=\textwidth]{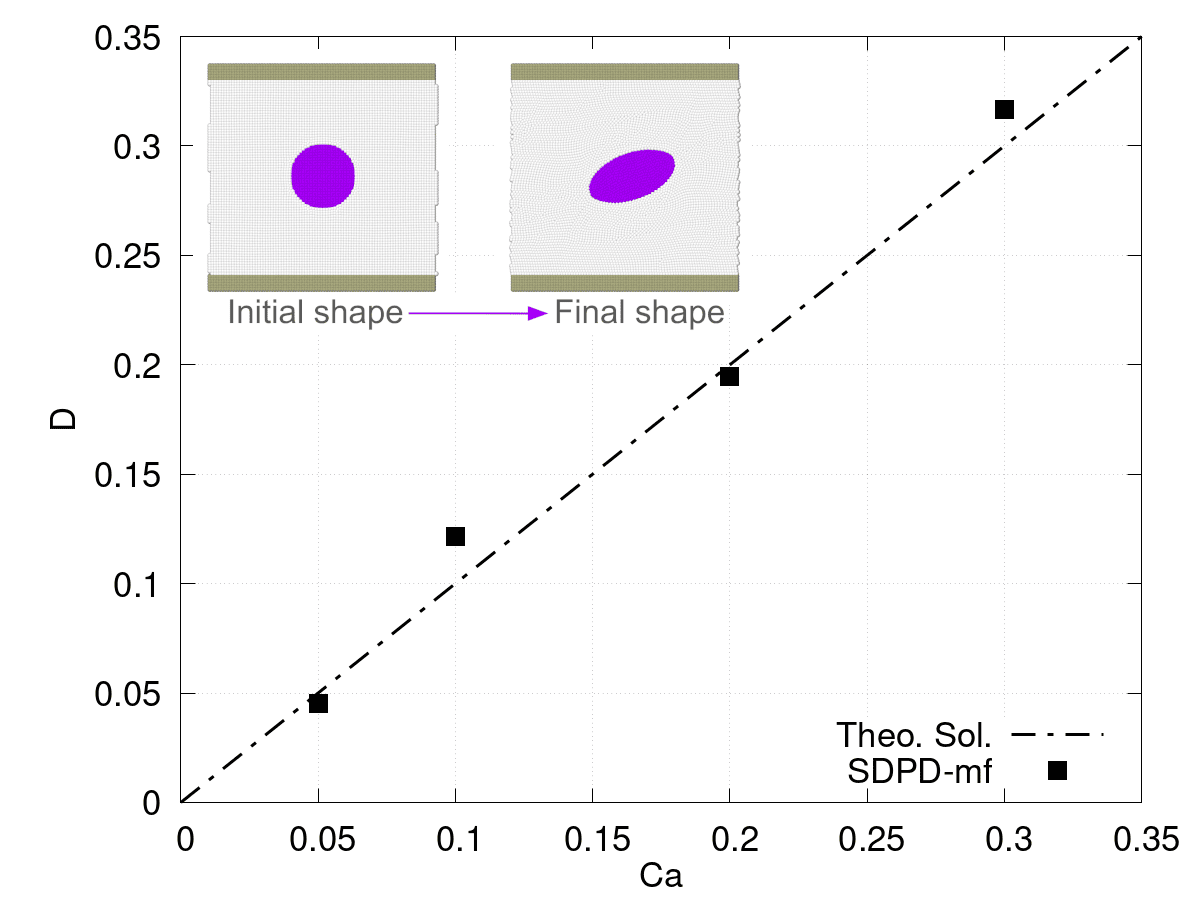}
         \caption{D vs Ca}
         \label{fig:2-b}
     \end{subfigure}
     \hspace{0.5cm}
     \begin{subfigure}[b]{0.22\textwidth}
         \centering
     \includegraphics[width=\textwidth]{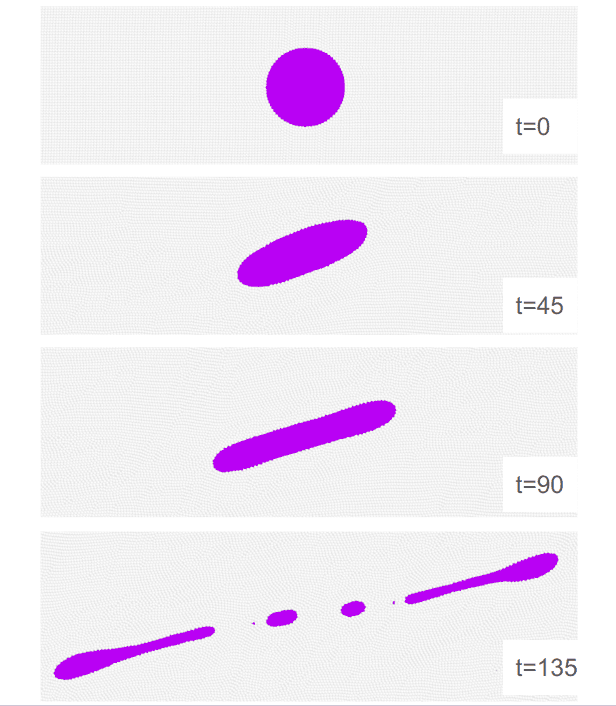}
         \caption{Droplet break-up}
         \label{fig:2-c}
     \end{subfigure}
\caption{Validation of the methodology for multiphase  Poiseuille flows using (a) a channel with two phases at different viscosity ratios, (b) Taylor deformation vs Capillary number for a droplet under a shear flow and (c) droplet break-up. }
\label{fig:2}
\end{figure*}

As discussed in the description of the numerical methodology, the interaction force (\Cref{eq:pairwise}) induces negative virial pressure in the system, which can lead to numerical instabilities. To avoid this, we introduce a background pressure $P_b$ (directly linked to $\sigma_0$) in \eqref{eq:pressure} that counterbalances the negative virial pressure. To verify that $P_b$ does not introduce artifacts, we initially evaluate single-phase flows under various configurations, then, we validate  dynamic conditions for multiphase systems in Poiseuille and shear flows. For single-phase flows, the system's behavior should align with the standard SDPD formulation \eqref{eq:deterministic}, regardless of the added pairwise force or background pressure. \ref{app1} presents a comparison of velocity and stress profiles for single-phase fluid flows against theoretical predictions and prior numerical results. The analysis demonstrates that incorporating $P_b$ not only avoids introducing artifacts into the simulation but is also essential for maintaining system stability. 

In the following, we validate the multiphase model by simulating two benchmark cases: a Poiseuille flow in a channel with two fluid phases and the dynamics of a droplet suspended in a liquid domain. The first case allows us to assess the model's ability to capture the velocity profile and interface stability under varying viscosity ratios, while the second case tests the model's capability to simulate droplet deformation and breakup under shear flow conditions. First, we consider a Poiseuille flow  in a channel containing two fluid phases, $1$ and $2$, initially distributed in the lower and upper half of the channel, respectively. We set a constant viscosity for the phase $1$ ($\eta_1 = 1$) and  different the viscosities for the phase $2$, by varying  viscosity ratio $\lambda = \eta_1 / \eta_2$ in the range $\{1, 2, 5, 10\}$. The height of the channel is $L/\Delta x=50$. For this setting, the theoretical velocity profile follows~\cite{tartakovsky2009lagrangian}  
\begin{equation}
v(x) = \rho g L^2 
\begin{cases} 
    - \frac{1}{2 \eta_2} \left( \frac{y}{L} \right)^2 + \bar{\eta}  \frac{1}{\eta_2} \frac{y}{L} & 0 < y< L/2, \\
    \frac{1}{2 \eta_1} \left( 1 - \left(\frac{y}{L} \right)^2 \right) + \bar{\eta}  \frac{1}{\eta_1} \left(\frac{y}{L} - 1 \right)  & L/2 \leq y < L,
\end{cases}
\end{equation}
where
\begin{equation}
\bar{\eta} = \left(  \frac{3}{4\eta_1} + \frac{1}{4\eta_2} \right) \left(  \frac{1}{\eta_2} + \frac{1}{\eta_1} \right)^{-1},
\end{equation}
In \Cref{fig:2-a}, we compare the velocity profiles simulated for the different viscosity ratios with the expected theoretical predictions. Notably, the model accurately capture the retraction in the parabolic profile as the viscosity ratio increases, preserving the stability of the interface between the two phases.  

Next, we validate the model by simulating the dynamics of a droplet ($d$) of radius $R_0$, suspended in a continuous fluid ($f$) phase under shear flow conditions. The droplet is initially spherical and is subjected to a shear flow induced by a constant velocity gradient $\dot{\gamma}$.  The system is confined in a square simulation box of side length $L/\Delta x=100$.

In \Cref{fig:2-b}, we present the measured Taylor deformation parameter $\mathcal{D}$ for different capillary numbers $Ca = [0.05 - 0.3]$. We consider a droplet of size $R_0/\Delta x = 12.5$, within two walls moving at velocity $U$ and $-U$, respectively, in a square channel of size $L = 6R_0$ . The physical properties of the droplet and fluid are $\rho_f=\rho_d=1$, $\eta_f=\eta_d=5$ and $\sigma=2.5$. We vary $Ca$ by changing the walls velocity, leading to Reynolds number in the range $\{0.025, 0.05, 0.075, 0.1\}$ always smaller than 1. In \Cref{fig:2-b}, we can observe an excellent agreement of our numerical results with the theoretical predictions \cite{pan2014dissipative} for $Ca<0.4$, where $\mathcal{D}$ can be approximated as $\mathcal{D} = (19 \lambda + 16)/(16 \lambda + 16) Ca$. 

In \Cref{fig:2-c}, we further explore the droplet dynamics as the capillary number increases, where the droplet undergoes breakup due to the unbalance between viscous forces and surface tension. \Cref{fig:2-c} shows the evolution from the undeformed state to the break-up of a droplet with $R_0/\Delta x=15$, $L=10 R_0$, $\sigma=1$, $\eta_f=\eta_d=1$ and $\gamma=0.3$ corresponding to a $Ca=0.9$.  The breakup occurs when the droplet's aspect ratio exceeds a critical value, leading to the formation of smaller droplets. The breakup process is characterized by the formation of a neck at the droplet's center, which eventually leads to the separation of the droplet into two smaller droplets. This behavior is consistent with previous studies on droplet dynamics in shear flows \cite{pan2014dissipative,wang2023dynamics,zhang2023shear}. 
\subsection{Thixotropic model validation}

\begin{figure*}[t!]
\centering
     \begin{subfigure}[b]{0.32\textwidth}
         \centering
     \includegraphics[width=\textwidth]{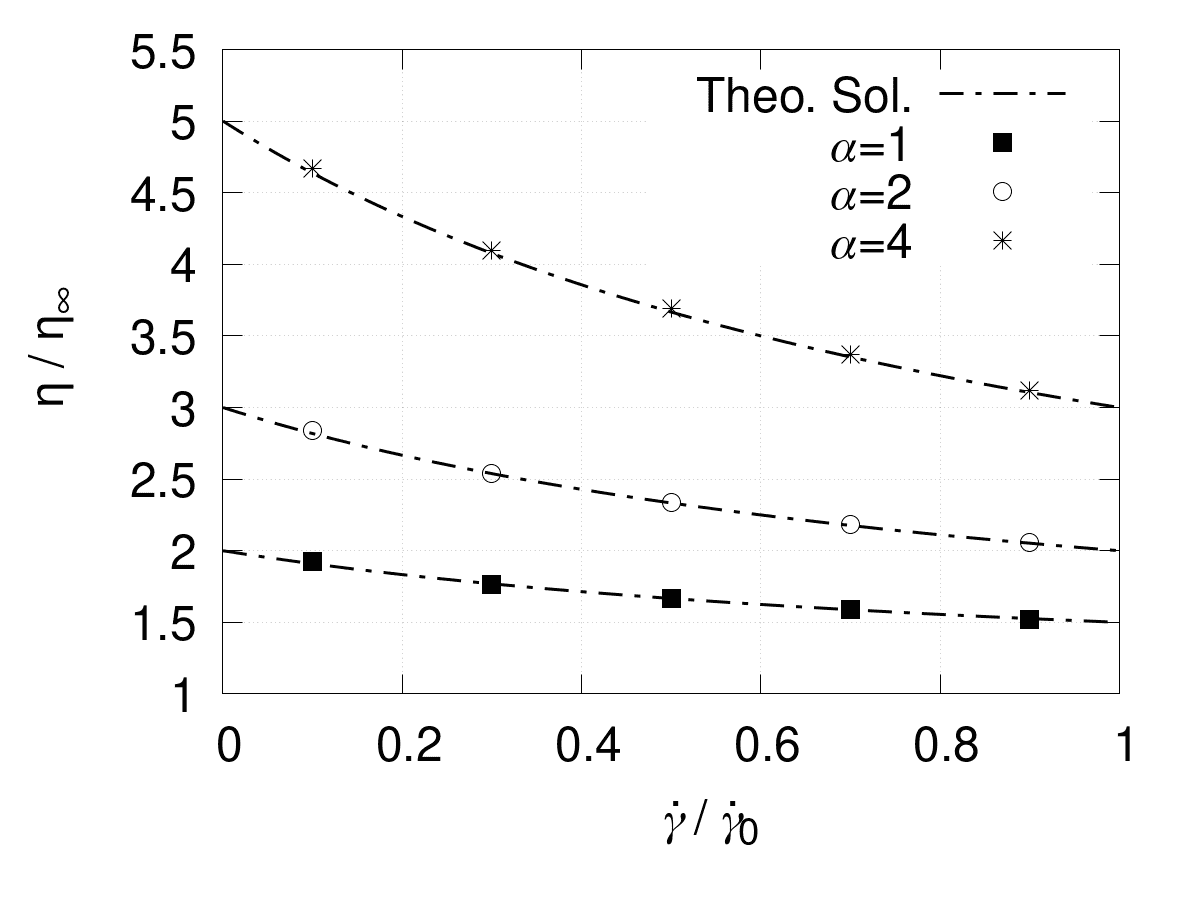}
         \caption{Steady shear viscosity}
         \label{fig:3-a}
     \end{subfigure}
     \begin{subfigure}[b]{0.32\textwidth}
         \centering
     \includegraphics[width=\textwidth]{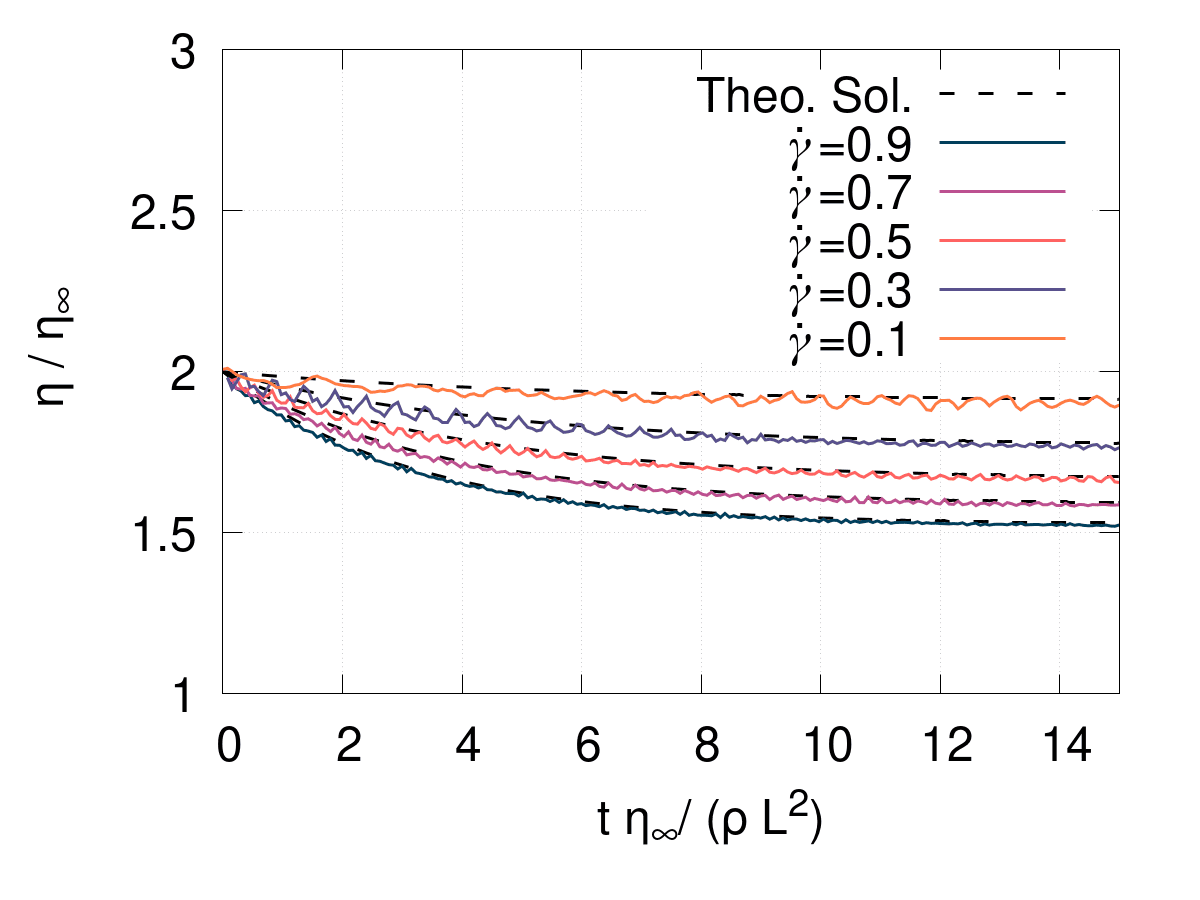}
         \caption{Viscosity evolution for $\alpha=1$}
         \label{fig:3-b}
     \end{subfigure}  
\caption{Thixotropic model validation for a simple shear flow for $\alpha$ range of (1,2,4)). Model parameters are based on the validation process developed by \cite{rossi2022sph}}
\label{fig:3}
\end{figure*}

We validate our implementation of the thixotropic model considering a single-phase fluid in a simple shear flow within two parallel plates moving in opposite directions.  We analyze five particular cases for which we set a sonic time scale $\tsnd= 0.5$, characteristic flow time scale $\tconv= \{10, 3.33, 2, 1.43, 1.11\}$ and thixotropic time $\tact=40$, following the hierarchy $\tsnd \ll \tconv, \tact$. We define a distance between plates $L/\Delta x = 50$ and a set of five velocities corresponding to $U=\{0.5, 1.5, 2.5, 3.5, 4.5\}$. The fluid properties are set as $f_0 = 1$, $\eta_\infty = 15$, $\rho = 1$, and $\ratfo = \ratde = 0.025$ (all within the adopted unit system), to fulfill the condition $\lambda_{\text{thix}} > L^2 \rho / \eta_\infty$~\cite{rossi2022sph}. These parameters result in a Reynolds number $Re = \{0.33, 1, 1.66, 2.33, 3\}$ and a Thixotropic number $Th = \{0.1, 0.3, 0.5, 0.7, 0.9\}$. Following Rossi et al. \cite{rossi2022sph}, we track the evolution of the steady state viscosity $\eta$ from the measured tangential force on the walls $F_x$, such that $\eta(\dot{\gamma}) = {F_x}/{A_w \dot{\gamma}}$, where $A_w$ is the wall area.  

In \Cref{fig:3}, we summarize the results of the thixotropic model validation, evidencing the accuracy of the model in both steady and transient conditions. In \Cref{fig:3-a} we present the variation in the measured steady state values of $\eta/\eta_{\infty}$ (for various values of $\alpha = [1, 2, 4]$ in \Cref{eq:etavar}), as the shear rate $\dot{\gamma}$ increases, following an excellent agreement with the theoretical model \Cref{thixo_1} (see \Cref{sec:met}). In \Cref{fig:3-b}, we further present the transient behavior of $\eta(t)$ for a fluid with $\alpha=1$ ($\eta_{\text{max}}/\eta_{\infty}=2$) under different shear rates. The results evidence the transient microstructure evolution, leading to larger microstructure destruction (lower viscosity) as the applied shear rate increase. For the lowest values of $\dot{\gamma}$, noticeable oscillations in the instantaneous force measurement lead to small fluctuation in the  estimated $\eta$. However, these numerical oscillations do not affect the overall trend of the viscosity evolution, which remains consistent with the expected behavior of a thixotropic fluid. 

This validation highlight the interplay between the characteristic time scales. The variation in the flow time scale $\tconv$ relative to the thixotropic relaxation time $\tact$ reveals that when $\tconv \ll \tact$, the fluid response is dominated by the slow restructuring of the microstructure, leading to significant transient viscosity effects. Conversely, for larger $\tconv$, the system would approach steady-state more quickly, and thixotropic effects would be less pronounced.

\section{Case Studies: Thixotropic Multiphase Dynamics from Phase Separation to Microfluidics}

We now demonstrate the model's versatility in capturing complex multiphase dynamics under different physical conditions. Specifically, we examine: (i) proteins liquid--liquid phase separation (LLPS), (ii) the behavior of droplet suspensions, (iii) droplet transport through a periodically constricted channel, and (iv) droplet coalescence within microfluidic devices. 

\subsection{Liquid-Liquid Phase Separation (LLPS) and aging}

Liquid--liquid phase separation leading to droplets formation is a wide-spread phenomena occurring in biological processes \cite{alberti2017phase,alberti2019considerations,feng2019formation}, such as the formation of organelles in cells \cite{volkov2024phase,courchaine2016droplet} and coacervates in neurodegenerative diseases \cite{Gracia2022,boyko2022tau,zbinden2020phase}. The process is characterized by the formation of solute-rich droplets dispersed in a solute-poor continuous phase, where the microstructural properties of the dispersed phase can change over time due to different physical process (coacervation, gelation, cristalization, etc), leading to a thixotropic behavior. In this context, the thixotropic multiphase model can be used to describe the dynamics of proteins LLPS, where phase separation time scales $\lambda_{\sigma}$ and thixotropic time scales $\lambda_{\text{thix}}$ are interwined. 

\begin{figure}[htbp!]
\centering
     \includegraphics[width=1.0\columnwidth]{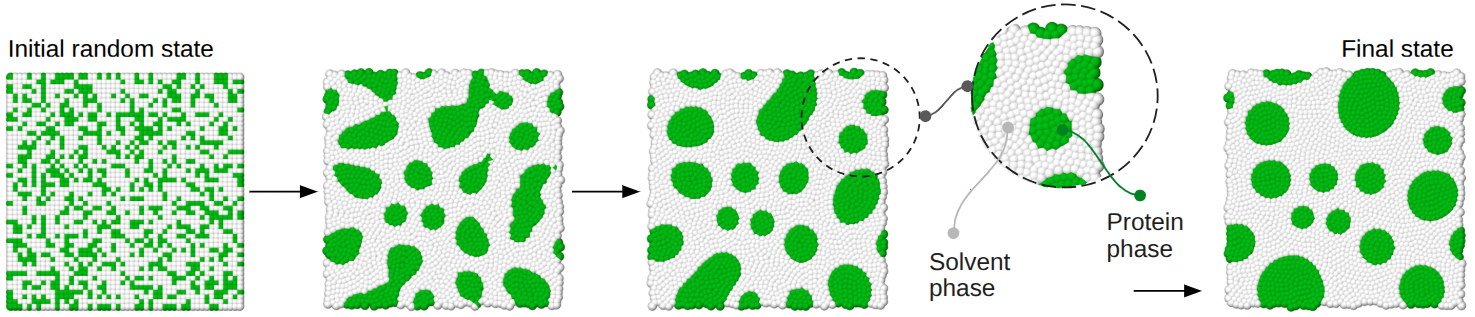} 
\caption{Phase separation evolution from the initial homogeneous state to the final phase-separated state.}
\label{fig:4}
\end{figure}

We investigate systems containing two type of particles, describing a disperse or "protein phase" ($p$) and a continuous or "solvent phase" ($s$), with volume fractions $\Phi_p$ and $\Phi_s = 1-\Phi_p$, respectively. As seen in \Cref{fig:4}, the system evolves from an initial homogeneously mixed condition to a final steady state, where the protein phase separate into droplets. Conceptually, we can think of the protein phase as a collection of protein aggregates dispersed in a continuous solvent phase. The protein phase is characterized by a critical concentration of proteins that triggers the phase separation process, leading to the formation of protein-rich droplets. The solvent phase contains the majority of the solvent and is responsible for the transport and interaction of the protein droplets. Note that we do not model explicitly the protein molecules but rather consider the protein phase as a thixotropic fluid which viscosity changes as protein aggregates form.

For a given $\Phi_p$, we can expected that a characteristic time scale $\lambda_{\sigma}$ for the phase separation process lies between a coarsening time scale $R^2/D_p$ and the ratio between the thixotropic and interfacial tension time scales $\tact/\tsig$. This is, $R^2/D_p < \lambda_{\sigma} < \tact/\tsig$, where $R$ is the mean radius of the droplets and $D_p \propto \kbt/(\eta_s h^2)$ is the diffusion coefficient of the protein phase. The coarsening time scale represents the time it takes for droplets to grow by diffusion, while the ratio $\tact/\tsig$ represents the relation between the microstructural restructuring and capillary-driven coalescence. The thixotropic time scale $\lambda_{\text{thix}}$, governs the rate at which the protein fluid evolves into highly viscous aggregates. The viscosity of the protein phase is assumed to increase with time, with an initial microstructural scalar parameter $f_0 = 0$, as the protein aggregates form and grow. Thus, $\eta_{\infty}$ is the initial  viscosity of the diluted protein phase. 

\begin{figure*}[t!]
    \centering
         \begin{subfigure}[b]{0.33\textwidth}
             \centering
         \includegraphics[width=\textwidth]{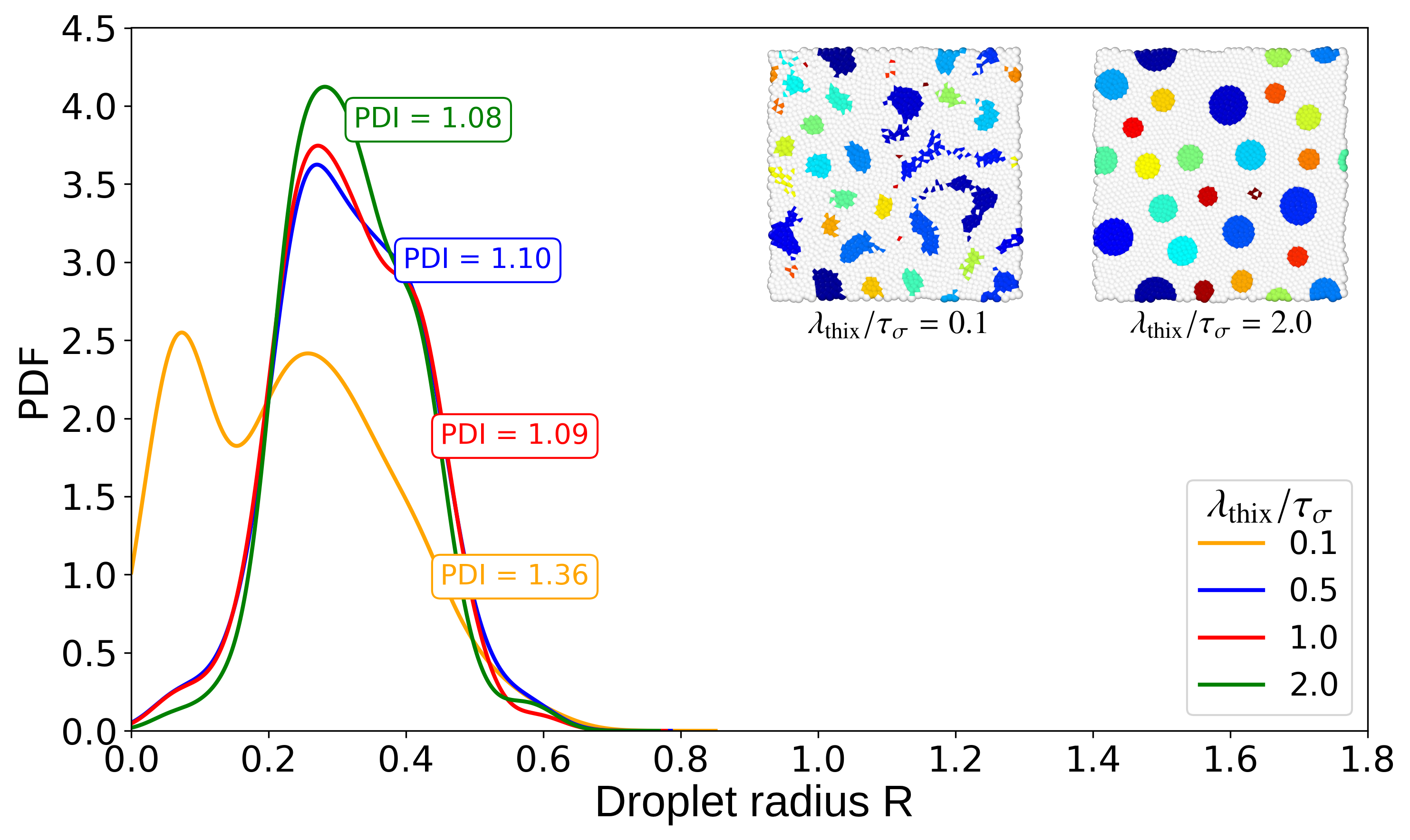}
             \caption{$\tact/\tsig$}
             \label{fig:5-a}
         \end{subfigure}
              \begin{subfigure}[b]{0.33\textwidth}
             \centering
         \includegraphics[width=\textwidth]{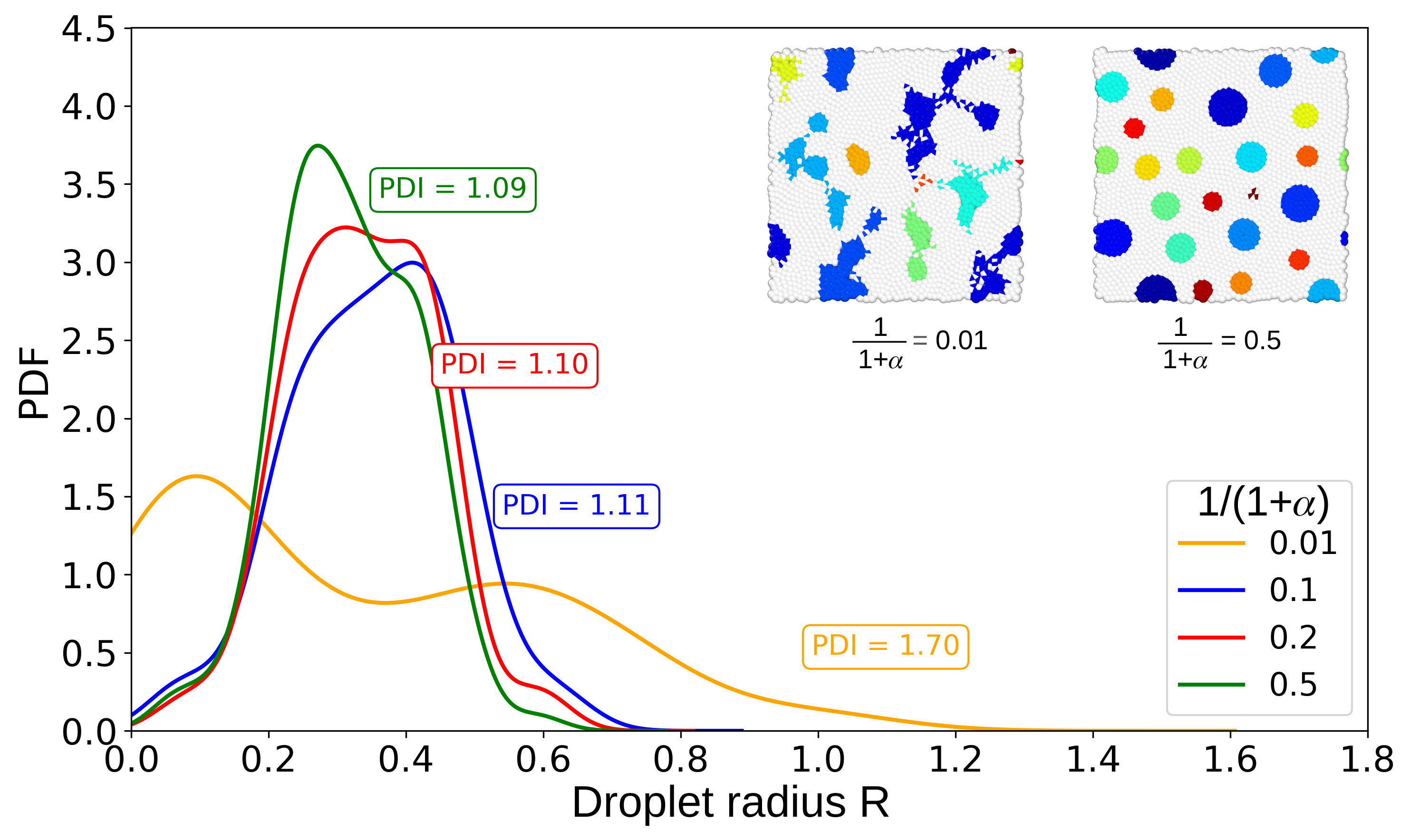}
             \caption{initial-to-final viscosity ${1}/{(1+\alpha})$} 
             \label{fig:5-b}
         \end{subfigure}  
         \begin{subfigure}[b]{0.33\textwidth}
             \centering
         \includegraphics[width=\textwidth]{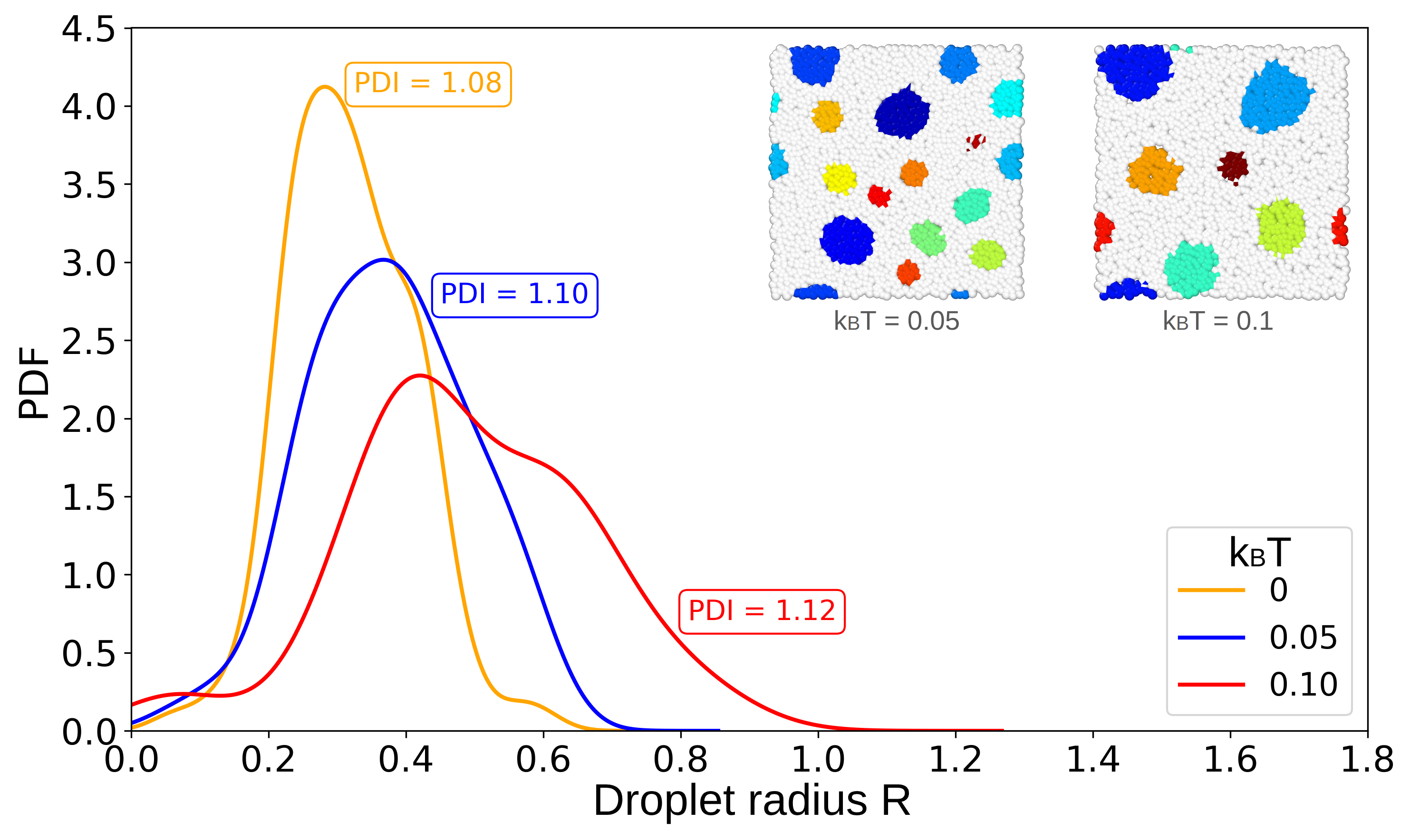}
             \caption{Thermal fluctuations $k_BT$}
             \label{fig:5-c}
         \end{subfigure} 
    \caption{KDE comparison for the different properties analyses including the PID value.  Inside snapshot illustrate the final configuration for two different system, where droplets are depicted in different colors for clarity. We employ the properties values for (a) $\tact/\tsig$ with $\sigma =[0.1, 0.5, 1.0, 2.0]$, $\alpha = 1$   (i.e. $\eta_{\text{max}}/\eta_{\infty}=2$), $k_BT=0$, for (b) initial-to-final viscosity with $\sigma =2.0$, $\alpha = [1,4,9,99]$, $k_BT=0$ and for (c) Thermal fluctuations with $\sigma =2.0$, $\alpha = 1$  (i.e. $\eta_{\text{max}}/\eta_{\infty}=2$), $k_BT = \{0, 0.05, 0.1\}$. For all simulations we use a fixed protein phase $\phi_p = 0.25$, viscosity of the solvent phase as $\eta_{s}=1$, thixotropic time scale $\lambda_{thix} = 1$ and $f_0=0$}
    \label{fig:5}
    \end{figure*}

To asses the interplay between the phase separation and thixotropic time scales, we investigate the effect of (i) variation of the ratio between the thixotropic characteristic time and interfacial tension time scale $\tact/\tsig =\{0.1, 0.5, 1, 2\}$, (ii) ratio initial-to-final viscosity ${1}/({1+\alpha}) = \{0.01, 0.05, 0.1, 0.5\}$, and (iii) thermal fluctuations time scale $ \tau_t = \{20, 28\}$ obtained by the variation in the thermal energy $k_BT=\{0.05, 0.1 \}$ . To calculate the characteristic times, we use a normalized radius droplet with unit length as the characteristic length, i.e., $R=1$ for the case of $\tsig$, and the length of the box $L=10$ for the case of $\tth$. For simplicity, we study these parameters under a fixed  protein phase concentration $\phi_p = 0.25$. Note that since protein phase concentration $\phi_p$ can also affect the phase separation dynamics in a non-linear fashion. In \ref{app2}, we briefly illustrate the effect of $\phi_p$, where lower concentrations favor smaller droplets with narrower distributions, whereas larger concentrations lead to larger droplets or spinodal-type separation of the phases. We conduct LLPS simulations in square domains of size $L/\Delta x=50$, in quiescent conditions ($\dot{\gamma} = 0$) such that no microstructural breaking is considered. We fix the viscosity of the solvent phase as $\eta_{s}=1$. All simulations reported in this subsection were conducted at a fixed thixotropic time scale $\lambda_{\text{thix}}=1$, thus, the variations observed correspond to changes in the relative magnitude of $\lambda_{\sigma}$ with respect to $\lambda_{\text{thix}}$.

We quantify the impact of these variations, by estimating both the number $N_{\text{drop}}$ and size $M_{\text{drop}}$ of the droplets formed up to a final time of $t= 25 \lambda_{thix}$, where "size" refers to the radius of the droplet calculated as $R = (\sqrt{N_{part}/d_{eq}}) / \pi$. For each type of test, we perform five independent simulations with different random seeds to ensure statistical robustness. To streamline the analysis and comparison between systems, we compute the average molecular weight (AM), the median (MED), the weighted-average molecular weight (WA), and polydispersity index (PDI) (see \ref{app2}). 

In \Cref{fig:5-a}, we summarize the Kernel Density Estimation (KDE) droplet size distributions for systems with varying $\tact/\tsig$, along with the snapshots of the final state for two different systems (central tendency statistics for $\tact/\tsig=0.1$ and $\tact/\tsig=2$ are illustrated in \ref{app2} in \Cref{fig:B14-c} and \Cref{fig:B14-d}, respectively). These results reveal that for $\tact/\tsig \leq 0.1$, the droplets exhibit irregular shapes, resulting in a broad and polydisperse size distribution (PID $\sim 1.4$) skewed toward smaller droplets. In contrast, for $\tact/\tsig \geq 0.5$, the phase separation promotes the formation of well-defined droplets. The increase in $\tact/\tsig$ beyond $0.5$ seems to not further alter the droplet size distribution, with a PID $\sim 1.1$. This behavior suggests that once the coalescence time scale increases (larger $\tact/\tsig$) than the characteristic thixotropic restructuring time ($\lambda_{\sigma} < \lambda_{\text{thix}}$), the dynamics of the system are governed primarily by thixotropic kinetics. Thus, further increases in $\tact/\tsig$ do not accelerate droplet coalescence, since the aging of the protein phase sets the effective upper limit for phase separation dynamics.

The variation of the ratio between the thixotropic and interfacial tension time scales, $\tact/\tsig$, shows the principal relation between microstructural restructuring and capillary-driven coalescence. When $\tact/\tsig \leq 0.1$, the interfacial dynamics evolve much slower than the microstructural relaxation, leading to incomplete coalescence events and irregular droplet morphologies. In this regime, extremely polydisperse and kinetically trapped configurations are produced as a result of the rapid interface formation that predominates before the material can reconstruct its internal structure. The two mechanisms operate over similar time scales as the ratio rises approaching $\tact/\tsig \approx 0.5$, enabling a more balanced evolution in which thixotropic rebuilding gradually stabilizes the droplets while surface tension efficiently reduces interfacial energy. Beyond this threshold ($\tact/\tsig \gtrsim 1$), the system reaches a quasi-steady regime in which droplet growth and coalescence are limited by the slow microstructural relaxation. In this limit, the thixotropic aging becomes the dominant factor controlling the overall kinetics of phase separation, and further increases in $\tact/\tsig$ do not significantly alter the droplet size distribution. 

In \Cref{fig:5-b}, we present the variations in the KDE curves for different values of the initial-to-final viscosity ratio $1/(1+\alpha)$. Representative histograms and central tendency statistics for $1/(1+\alpha) = 0.01$ and $1/(1+\alpha) = 0.5$ are also shown in \ref{app2} (\Cref{fig:B14-e} and \Cref{fig:B14-f}, respectively). The results indicate that lower initial-to-final viscosity ratios $1/(1+\alpha)\leq0.01$ lead to poorly developed structures with a larger molecular weight and broader distribution (PID $=1.7$). In contrast, as $1/(1+\alpha)$ increases the polydispersity of the systems reduces, favoring narrower size distributions. At the fixed $\lambda_{\text{thix}}=1$ used, a lower initial-to-final viscosity ratio implies that the change in viscosity of the protein phase is significant as the microstructure changes, leading to a very fast reduction in the capillary velocity, and consequently shifting the phase separation time scales towards $\lambda_{\text{thix}} \ll \lambda_{\sigma}$. As a result, the propensity of the system to phase separate decreases rapidly with the system aging, leading to poorly developed polydisperse structures. Conversely, when $1/(1+\alpha)\gtrsim 0.1$, the relative change in viscosity is less pronounced, so that $\tact/\tsig$ remains sufficiently high during phase separation.

Finally, in \Cref{fig:5-c}, we present the effect of thermal fluctuations over the droplets size distributions (see also \ref{fig:B14-g} and \Cref{fig:B14-h} for representative central tendencies for $k_BT = 0.05$ and $k_BT = 0.1$, respectively), for systems with the protein phase characterized by $\lambda_{\text{thix}}=1$, $\beta=1$, $\alpha = 1$ (i.e.$\eta_{\text{max}}/\eta_{\infty}=2$), and $\sigma_0=2$ (leading to $0.5 \leq \tact/\tsig \leq 1$). On the range of $\tth$ (and $\tact/\tsig$) explored, the thermal fluctuations are in the same scale than interfacial tension effects to ensure the phases remain immiscible. However, the diffusive transport of the SDPD particles allow us to enhance the coarsening of the droplets ($R^2/D_p$), leading to an overall faster phase-separation. For $k_BT=0.1$ for instance, a complete phase separation (single droplet) occurs over a time span $t=45 \lambda_{\text{thix}})$. Therefore, to facilitate the analysis at different $\tth$, we define a fixed final time $t=25 \lambda_{\text{thix}}$ to compare the different conditions. 

In the absence of thermal fluctuations, we have that $R^2/D_p \gg 1$ and the system yields relatively small, well-defined droplets with uniform size distributions, with droplet growth controlled mainly by $\tact/\tsig$. The resulting distribution remains relatively uniform, but coarsening is incomplete within the simulated time. Decreasing $\tth$ (i.e. increasing the $k_BT$), enhances diffusive transport, thereby reducing the coarsening time scale $\lambda_{\sigma}\sim R^2/D_p$. In \Cref{fig:5-c}, we can observe that the mean size and distribution of the droplets consistently increases with $k_BT$. However, the polydispersity index do not change significantly. When $\lambda_{\sigma}$ becomes shorter than $\lambda_{\text{thix}}$, droplets coalesce more rapidly leading to larger droplets, and the dynamics variation in $\tact/\tsig$ (due to the change  in viscosity) do not dominate the phase separation. Thermal fluctuations can dynamically modulate the interplay between $\lambda_{\sigma}$ and $\lambda_{\text{thix}}$. At high $\tth$ (low $k_BT$) the dynamics are limited by thixotropy, whereas as the coarsening of the phases is favored,  the phase separation progresses rapidly enough to control the system’s evolution.

The variation of the thermal time scale $\tth$ modulates the balance between diffusive transport, interfacial dynamics, and thixotropic restructuring. For large $\tth$ (low $k_BT$), phase separation is mainly governed by $\tact$ and $\tsig$, yielding slow coarsening and uniform droplets. When $\tth$ is reduced, enhanced diffusion accelerates droplet coalescence, shortening $\lambda_{\sigma}$ and facilitating more rapid droplet growth. Once $\lambda_{\sigma}$ becomes comparable to or smaller than $\lambda_{\text{thix}}$, the influence of thixotropy weakens and thermal diffusion governs the phase separation. Consequently, $\tth$ sets the transition between a regime governed by thixotropy and one controlled by diffusion. 

\subsection{Thixotropic flows in channels: Emulsions and suspensions}

We now extend the multiphase thixotropic model to describe complex emulsions and suspensions. These systems are of broad relevance since thixotropy has been extensively reported in a several contexts. For example, biological suspensions such as blood and mucus exhibit pronounced time-dependent viscosity, where microstructural rearrangements under flow lead to reversible decreases in viscosity followed by recovery at rest \cite{baskurt2003,lai2009}. Similarly, a wide range of synthetic materials—including concentrated emulsions, colloidal dispersions, and clay-based suspensions—have been shown to display aging, shear-induced breakdown, and subsequent structural rebuilding, all of which give rise to characteristic thixotropic rheology \cite{barnes1997,mewis2009}. 

\begin{figure}[hbtp!]
\centering
     \includegraphics[width=0.48\textwidth]{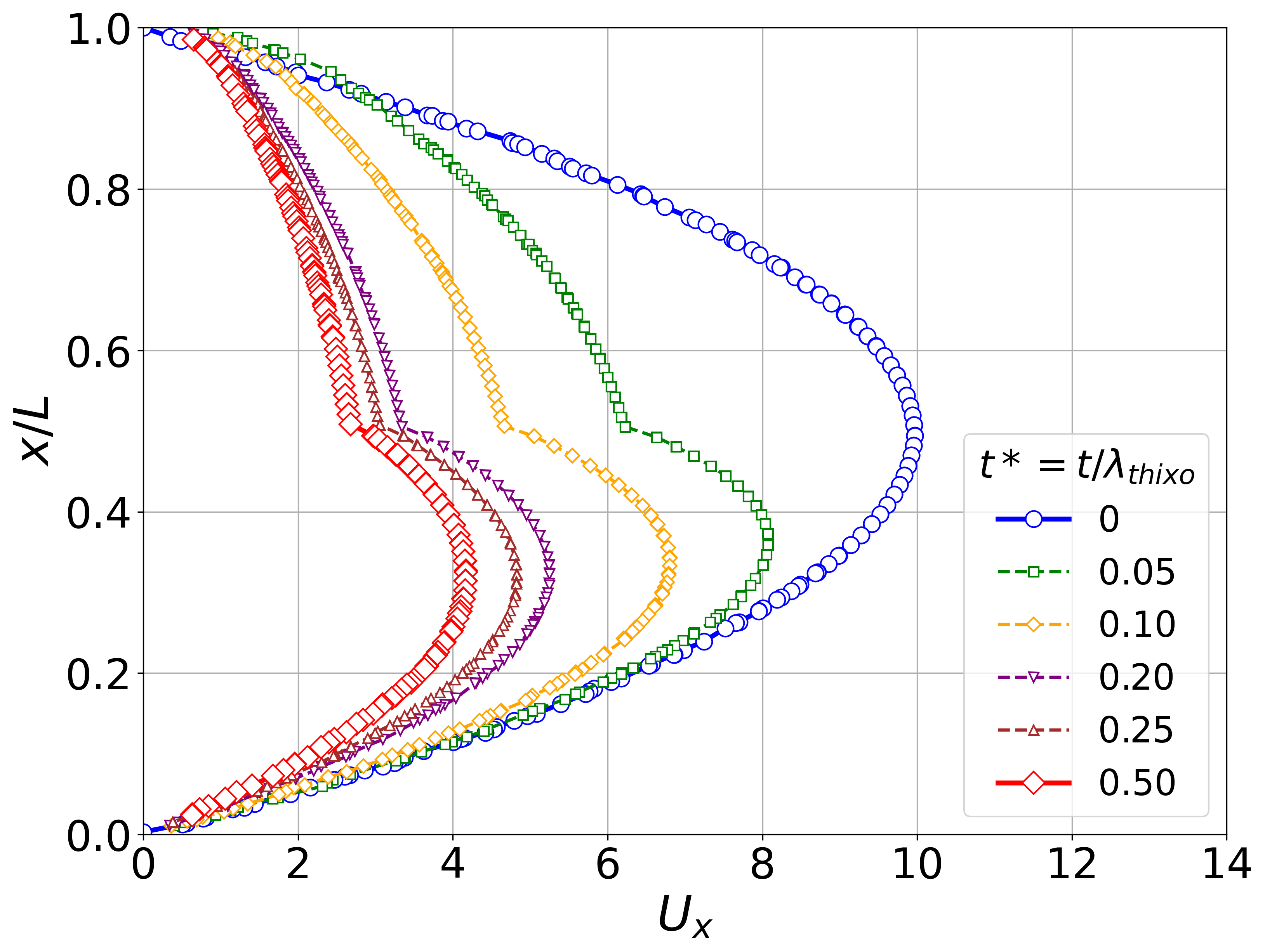}
\caption{Velocity profiles for a two phases flow}
\label{fig:6}
\end{figure}

We begin by considering the simplest configuration of a fully separated two-phase system (phases $\phi_A$ and $\phi_B$) under Poiseuille flow. To avoid spurious artifacts from numerical stabilization, the simulations are initialized with the corresponding stabilized Newtonian velocity profile. In this setup, phase $\phi_A$ maintains Newtonian behavior throughout the simulation, while phase $\phi_B$ evolves according to the thixotropic model. We employ a square channel with a characteristic length of $L/\Delta x = 50$ and a maximum velocity of $U = 10$, resulting in a characteristic flow time scale of $\tconv = 1$. Also we set a sonic time scale $\tsnd= 0.2$ and thixotropic time $\tact=100$ follow the hierarchy $\tsnd \ll \tconv, \tact$. We analyze the effects of microstructure destruction to formation rates by setting $\beta = 1$, the constitutive parameter $\alpha = 4$, and  with initially formed ($f_0 = 1$) microstructure. This selection of parameters results in a Reynolds number $Re=1$ and a thixotropic number $Th=1$. For reference, we also perform a simulation in which $\phi_B$ remains Newtonian but with its viscosity fixed at the maximum value prescribed by the thixotropic law. As illustrated in \Cref{fig:6}, the velocity profile in the thixotropic case gradually evolves from the Newtonian baseline toward the limiting profile corresponding to the maximum viscosity. This evolution reflects the interplay of time scales because the convective time $\tconv$ is much shorter than the thixotropic relaxation time $\tact$. The microstructure does not fully rebuild during the initial advection of the fluid, resulting in a transient velocity profile. Over longer times, the microstructure reaches its equilibrium, increases the effective viscosity, and stabilizes the velocity profile. These results demonstrate that the model accurately captures the emergence and stabilization of interfaces in multiphase Poiseuille flow under thixotropic conditions.

Now, we proceed to explore emulsions of \textit{i)} Newtonian droplets in thixotropic continuous phase, and \textit{ii)} thixotropic droplets in a Newtonian continuous phase. To avoid the merging of the droplets during flow, we introduce an additional pair-wise soft repulsive force $f_{\text{drop-drop}}= \epsilon_{\text{drop}} (1 -r/r_c)^2 $, between particles belonging to different droplets, where $r_{c}=4 \Delta x$ is the range of the interaction, and $\epsilon_{\text{drop}}$ is the strength of the repulsion that we normalize by a characteristic capillary force scale $\sigma_0 / r_c$, so that $\tilde{\epsilon}_{\text{drop}} = \epsilon_{\text{drop}} r_c / \sigma_0$. In the present simulations, we set $\epsilon_{\text{drop}} = 2.0$, which corresponds to $\tilde{\epsilon}_{\text{drop}} \approx 8$. While this force ensures droplets stability do not introduce any additional artifact in our simulations, as we illustrate in \ref{app3} for fully Newtonian emulsions, where we show that the model consistently reproduces variations of the emulsion's viscosity with the droplet concentration, as theoretically predicted by Taylor \cite{taylor1932viscosity} for deformable particles.

First, we consider an emulsion composed of Newtonian droplets  ($\eta' = 5$) suspended in a thixotropic continuous phase. The system is subjected to a simple shear flow ($\dot{\gamma} = 0.5$) in a channel of length $L/\Delta x = 50$,  resulting in a characteristic flow time scale of $\tau_{\text{flow}} = 2$. The hierarchy of relevant time scales satisfies $\tsnd \ll \tvis, \tsig \ll \tconv, \tact$, where the sonic, viscous, interfacial, and thixotropic times are set to $\tsnd = 0.022$, $\tvis = 0.04$, $\tsig = 0.2$, and $\tact = 100$, respectively. To guarantee a scale separation between the long microstructural time scale and the short viscous ones \cite{rossi2022sph}, we set the parameters of the thixotropic model such that $\tact > L^2 \rho / \eta_\infty$. The continuous phase viscosity is set to $\eta_\infty = 5$, with $\beta = 1$, constitutive parameter $\alpha = 4$, and an initially formed microstructure ($f_0 = 1$).  The radius of the droplets is calculated as $R/\Delta x = (\sqrt{N_{part}/d_{eq}}) / \pi = 2.2$. The chosen parameters result in a Reynolds number $Re = 5$ and a Thixotropic number $Th = 0.5$. The systems are initialized as Newtonian emulsion, until the steady state of their velocity profile is reached. Then the thixotropic model is used for the droplet fluid. 

\begin{figure}[hbt!]
\centering
     \includegraphics[width=0.48\textwidth]{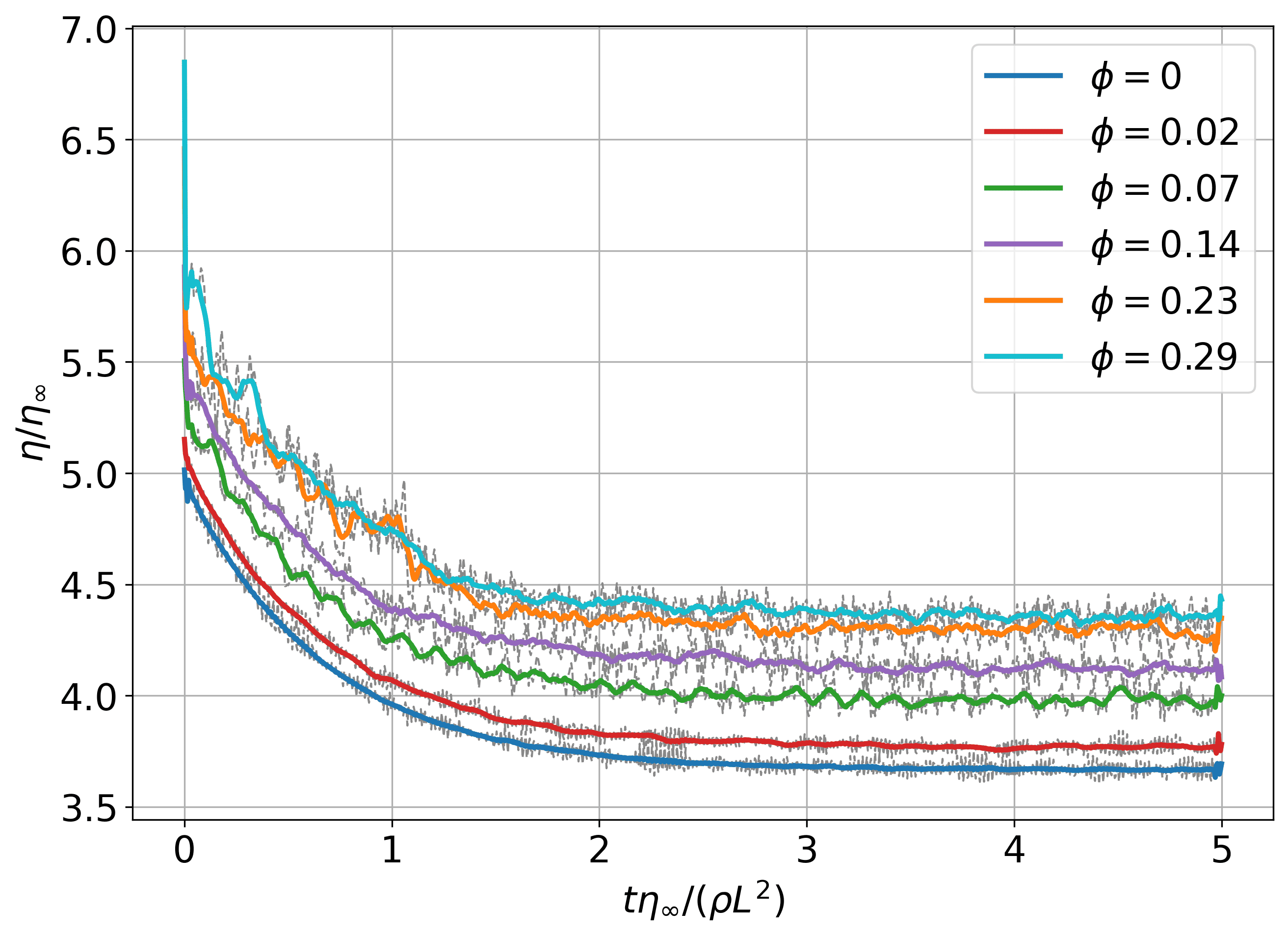}
\caption{Transient viscosity evolution in a channel with Thixotropic continuous phase for different values of $\phi$}
\label{fig:7}
\end{figure}

We test five volume fractions $\phi$ and monitor the emulsion behavior over time.  This volume fraction is calculated as the volume of the dispersed phase ($V_{dp}$) relative to the volume of the continuous phase ($V_{cp}$) i.e. $\phi = V_{dp}/V_{cp}$. In \Cref{fig:7}, we present the response in the measured viscosity, showing an initial drop from a peak effective viscosity to a lower asymptotic value as the microstructure relaxes from its initially formed state ($f_0=1$). This decay reflects the progressive breakdown of the microstructure and the transition of the system toward its thixotropic equilibrium. As expected, higher droplet concentrations lead to higher effective viscosities due to enhanced hydrodynamic interactions. Overall, the plot demonstrates that thixotropy not only reduces or increase (depends of $f_0$) the steady-state viscosity but also introduces a transient regime where the effective viscosity evolves dynamically, underlining the role of the thixotropic timescale in determining the apparent rheological response of the system. The observed transient behavior of the effective viscosity can be directly interpreted in terms of the hierarchy of time scales. The droplets are transported along the channel while largely preserving their original shape because the convective time scale $\tconv$ is larger than the interfacial time $\tsig$. The longest characteristic time, the thixotropic relaxation time $\tact$, governs how rapidly the microstructure responds to flow-induced stresses. Given that $\tact \gg \tconv$ and that the system starts from a fully structured state ($f_0 = 1$), the evolution is dominated by microstructure breakdown rather than rebuilding. Consequently, the effective viscosity decreases dynamically as the initially structured droplets progressively adapt to the local flow conditions.

In the case of emulsions of thixotropic droplets suspended on a Newtonian fluid, we simulate a Poiseuille flow in a domain of size $L/\Delta x= 100$ containing $25$ initially spherical droplets.  In a steady state, the fluid reaches a maximum velocity of $U=1.25$, resulting in a characteristic flow time scale of $\tconv = 16$. Although the specific value of $\tconv$ differs from the previous configuration, the relative ordering of the characteristic times remains unchanged, preserving the hierarchy $\tsnd \ll \tvis, \tsig \ll \tconv, \tact$. The continuous phase has a viscosity $\eta_s = 5$ and density $\rho = 1$, while the dispersed phase is characterized by $\eta_\infty = 5$, $\alpha = 4$ and $\beta = 10$. Both, initially fully-formed microstructure ($f_0 = 1$) and unformed  ($f_0 = 0$) are considered. As previously calculated, the radius of the droplets is equivalent to $R/\Delta x= 2.2$ . Again, we set the parameters of the model to satisfy the condition $\tact > L^2 \rho / \eta_\infty$  \cite{rossi2022sph}. This set of parameters yields a Reynolds number of $Re = 5$ and a Thixotropic number of $Th = 0.625$.

\begin{figure*}[hbtp!]
\centering
     \includegraphics[width=0.8\textwidth]{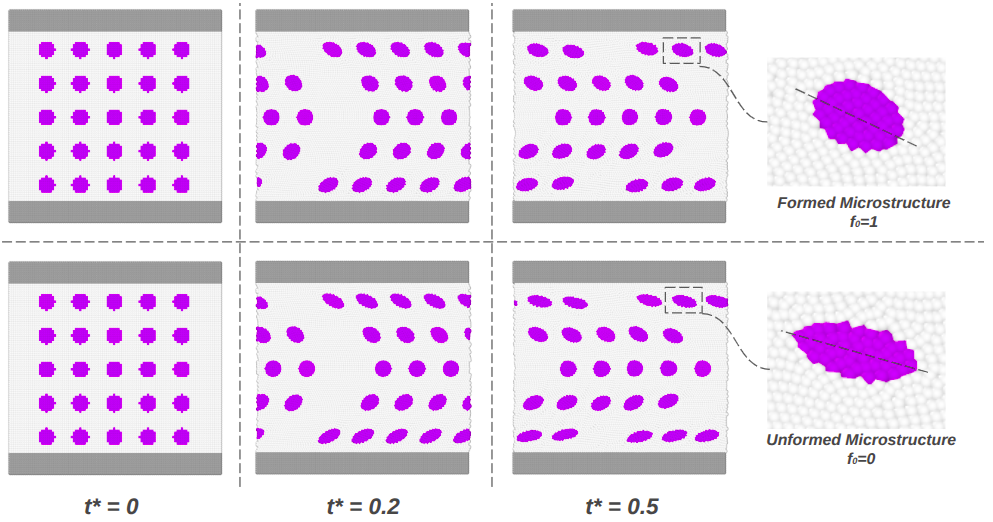}
\caption{Emulsion with Newtonian continuous phase and thixotropic dispersed phase. Comparison starting with the formed microstructure (top row) versus the unformed microstructure (bottom row). Here, $t^*=t/\lambda_{thix}$. The detail of the droplets is shown at $t^*=0.5$ (before reaching a steady state) to illustrate the differences in the transient state.}
\label{fig:8}
\end{figure*}

\Cref{fig:8} compares the transient evolution of systems starting with a formed microstructure (top row) and an unformed microstructure (bottom row). The droplet deformation depends on the local shear rate, being minimal at the channel center and increasing near the walls. Thus, the morphological evolution of the droplets in the middle of the channel is practically unaffected by the microstructure destruction, their morphology remains constant regardless of the initial $f_0$. As, we move away from the center, in the case of $f_0 = 1$, the initially high viscosity $\eta = \eta_\infty(1+\alpha)$ limits deformation and  $\lambda_{\text{thix}}$ governs the droplet response, with modest  droplet deformation across the domain. In contrast, droplets with $f_0 = 0$, starting with the base viscosity $\eta_\infty$, tend to deform more strongly under flow leading to significantly more elongated droplets near the walls. The behavior observed in the emulsion of thixotropic droplets in a Newtonian continuous phase can be interpreted similarly to the previously discussed  case.  As in the previous scenario the thixotropic relaxation governs the evolution of the microstructure. Consistently, droplets traveling in the center of the channel retain their morphology regardless of the initial state $f_0$, whereas those closer to the walls experience more pronounced deformation due to larger Ca numbers. We must note that in the steady state, both systems ($f_0=0$ and $f_0=1$) should reach an equivalent state of droplets deformation. However, their transient evolution evidence the interplay between flow-induced stress and microstructural response, and highlight potential emergence of heterogeneous droplets morphology. 

\subsection{Droplet dynamics in a periodically constricted channel}

We now analyze the transition of a thixotropic droplet through a periodically constricted channel. Understanding the dynamics of thixotropic fluids in this setting is particularly relevant in systems where fluids experience recurrent deformation and relaxation, including blood flow in the microcirculation, polymeric suspensions in microfluidics, and drilling muds in porous media. The constricted geometry provides a controlled framework for examining the coupling between microstructural dynamics and macroscopic transport properties.

\begin{figure*}[hbtp!]
\centering
     \begin{subfigure}[b]{0.7\textwidth}
         \centering
     \includegraphics[width=\textwidth]{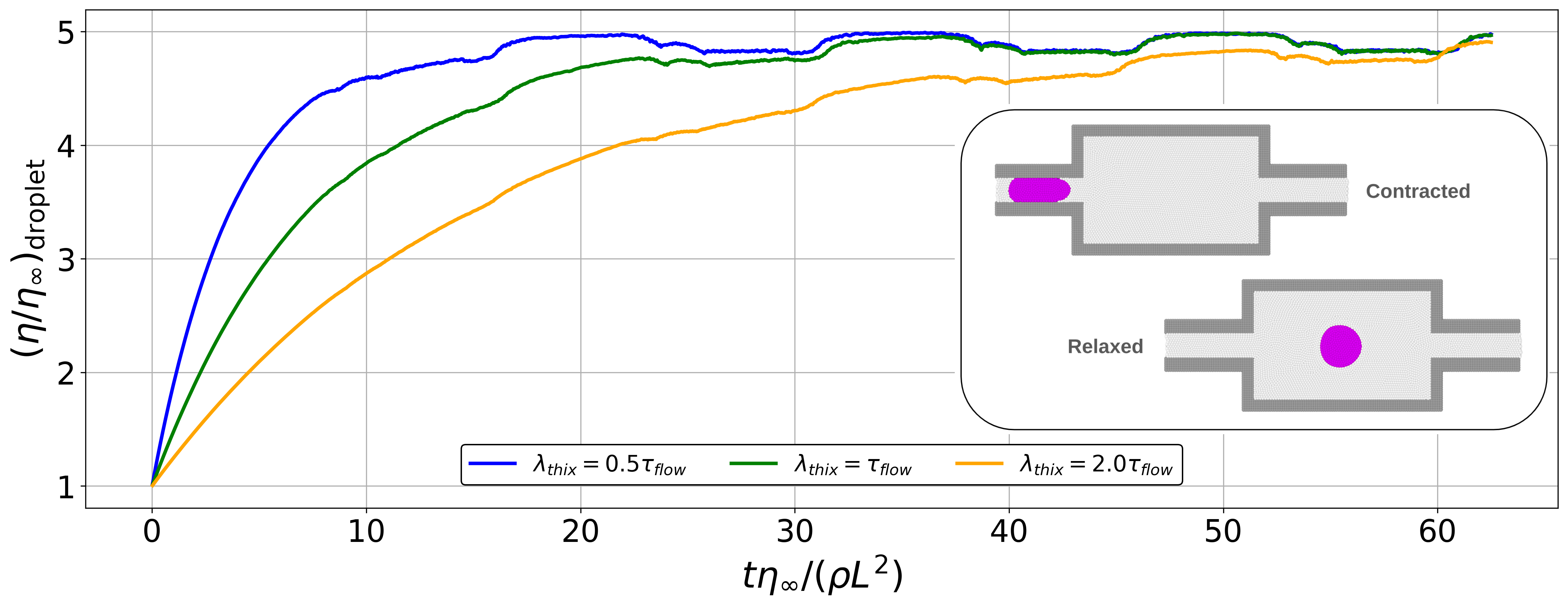}
         \caption{Initially unformed microstructure, ($f_0=0$)}
         \label{fig:9-a}
     \end{subfigure} 
     \begin{subfigure}[b]{0.7\textwidth}
         \centering
     \includegraphics[width=\textwidth]{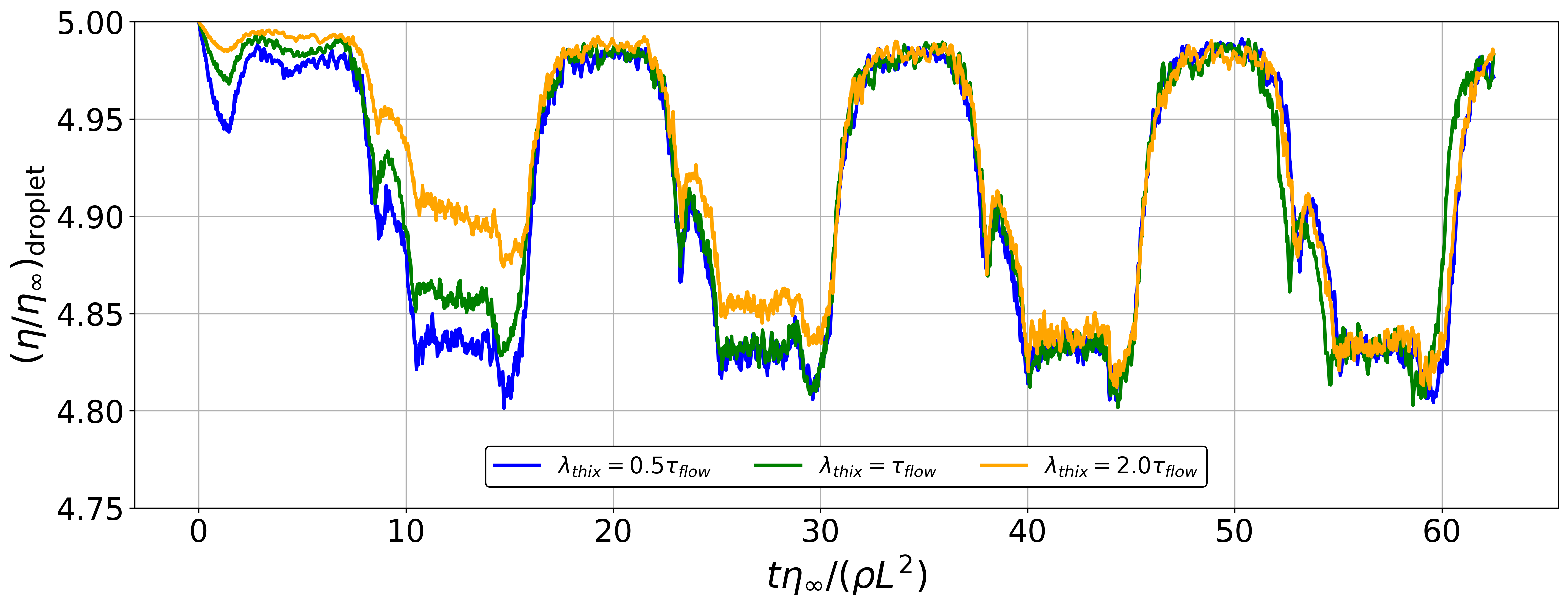}
         \caption{Initially fully formed microstructure ($f_0=1$)}
         \label{fig:9-b}
     \end{subfigure}
\caption{Transient evolution of the viscosity of a thixotropic droplet passing through a chamber with diameter change. Comparison at different values of $\lambda_{\text{thix}}$ at $\beta=1$ for cases starting with a) initially unformed microstructure $f_0=0$ and b) initially fully formed microstructure $f_0=1$.}  
\label{fig:9}
\end{figure*}

We consider a spherical droplet of size $R/\Delta x=10$ in a periodic channel of length $L_{cr} /\Delta x = 160$, constituted by two sections of length $L/\Delta x=(L_{cr}/(\Delta x))/2=80$, with heights $H_c/\Delta x= 11$ and $H_r/ \Delta x =50$ for the constrained and relaxed sections, respectively (see \Cref{fig:9-a}). The maximum flow velocity in the channel is $U = 0.017$. The relative lengths of the constriction and expansion determine whether the imposed shear aligns with the fluid's thixotropic timescales of breakdown and recovery. Short constrictions may not fully degrade the structure, while longer ones drive the fluid toward its thinned state. Likewise, the expansion length controls how much recovery occurs before the next constriction. This coupling dictates local viscosity profiles and impacts transport efficiency and flow stability. We define the simulations parameters to ensure proper timescales hierarchy ${\tsnd} \ << \ {\tvis} \ << \tconv,\tact $. Here, $\tsnd = 0.8$ and the viscous relaxation time ${\tvis}=20$. The maximum velocity reached by the fluid in the relaxed zone leads to $\tconv=200$. To investigate the interplay between flowing and thixotropic time scales, we explore three specific conditions, $\tact = 0.5\tconv$, $\tact = \tconv$, and $\tact =  2.0\tconv$. This corresponds to a Reynolds number of $Re = 0.17$ and a Thixotropic number $Th = 0.005$. Additionally, we analyze the effects of microstructure destruction to formation rates by varying the parameter $\beta = \{0.1, 1, 10, 100\}$. These experiments were conducted by by setting $\tact = \tconv$, resulting in $Re = 0.17$ and $Th = \{ 0.0005, 0.005, 0.05, 0.5 \}$. For the different scenarios we consider droplets with initially unformed ($f_0 = 0$) and formed ($f_0 = 1$) microstructure. We consider the transporting fluid to be Newtonian with viscosity $\eta = 1$, whereas the droplet has $\eta_\infty = 10$ and $\alpha = 4$ ($\eta_{\text{max}}=5\eta_\infty$). The density of both the fluid and the droplet is $\rho = 1$.

In \Cref{fig:9}, we provide the results for the droplet's viscosity evolution for different thixotropic time scales, at a fixed destruction-to-formation rate $\beta =1$. For the case of initially unformed microstructure $f_0=0$ (see \Cref{fig:9-a}), it can be seen how the viscosity starts at its limiting value ($\eta/\eta_\infty=1$) and evolves toward its maximum. As expected, for shorter $\lambda_{\text{thix}}$, the stabilization of the viscosity towards $\eta_{\text{max}}$ occurs in a fewer constraint-to-relaxed cycles within the channel, whereas for the largest $\lambda_{\text{thix}}$, the full-microstructure formation is not even reached after seven cycles on the domain. In contrast, in droplets with initially formed microstructure $f_0=1$ (see \Cref{fig:9-b}), the viscosity profiles for the different $\lambda_{\text{thix}}$, readily converge within two cycles in the domain.

\begin{figure*}[hbt!]
\centering
     \begin{subfigure}[b]{0.7\textwidth}
         \centering
     \includegraphics[width=\textwidth]{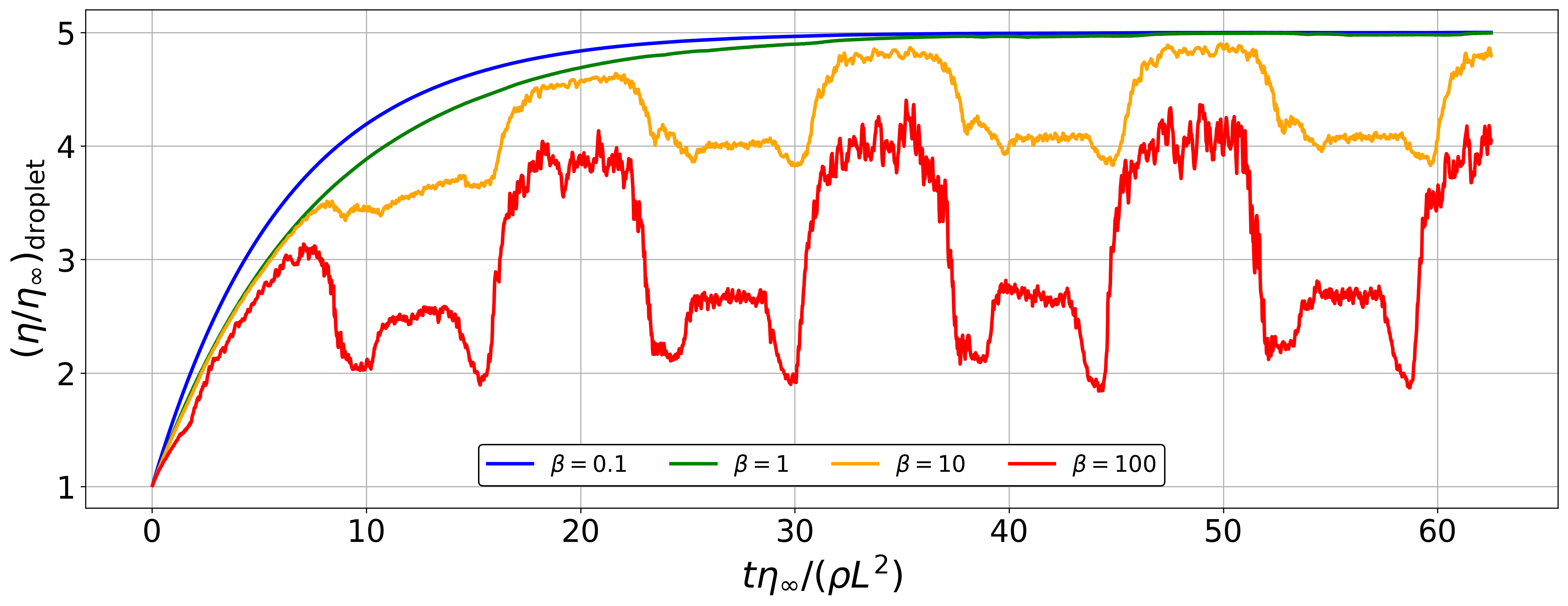}
         \caption{Initially unformed microstructure, ($f_0=0$)}
         \label{fig:10-a}
     \end{subfigure} 
     \begin{subfigure}[b]{0.7\textwidth}
         \centering
     \includegraphics[width=\textwidth]{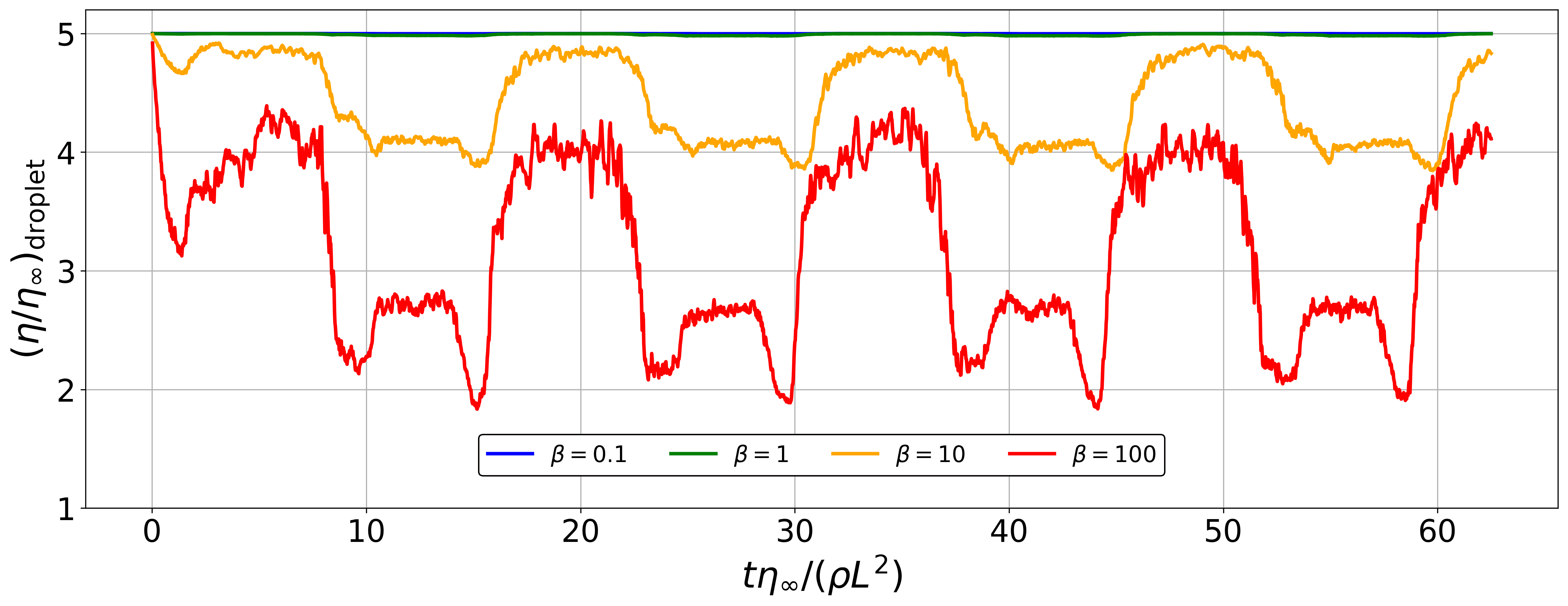}
         \caption{Initially fully formed microstructure ($f_0=1$)}
         \label{fig:10-b}
     \end{subfigure}
\caption{Transient evolution of the viscosity of a thixotropic droplet passing through a chamber with diameter change. Comparison at different values of $\beta$ at $\lambda_{\text{thix}}=200$ for cases with a) initially unformed microstructure $f_0=0$ and b) initially fully formed microstructure $f_0=1$.}
\label{fig:10}
\end{figure*}

For lower value of $\beta=1$ explored, the destruction of the microstructure do not play a major role during the early stages of the droplet flow. However, as $\beta$ increases, it is expected that larger viscosity variation can emerge as the abrupt change in shear rate between regions occurs. In \Cref{fig:10}, we compile the results of the viscosity variation for droplets with fixed thixotropic time scale to $\lambda_{\text{thix}} = \tau_{\text{flow}}$, for different values of $\beta$ and $f_0$. Consistently, we observe that the larger rates of microstructural destruction amplify the variations in shear rate, and consequently, the effective viscosity of the droplet. 

In general, the observed interplay suggests that by tuning fluid properties together with channel dimensions, devices can be specifically designed to either minimize viscosity fluctuations for stable transport or exploit them for enhanced mixing, separation, or controlled delivery. For instance, lab-on-a-chip platforms can exploit viscosity variations to enhance mixing efficiency in low-Reynolds-number regimes, while separation devices can leverage controlled differences in viscosity to direct particles or cells along distinct trajectories. This ability to suppress or amplify viscosity fluctuations through the combined tuning of fluid properties and channel geometry underscores a powerful design principle for engineering application-specific fluid behaviors.

\subsection{Droplet merging using micro-devices}

Within microfluidic analysis, the conditions and parameters governing droplet merging of great interest \cite{niu2008pillar,basu2013droplet}. Experimental techniques such as Droplet Morphometry and Velocimetry~\cite{basu2013droplet} (DMV) allow real-time observations of merging events on microscales resolutions. Moreover, microfluidic devices equipped with passive structures configurations~\cite{niu2008pillar} have been also successfully designed control contact-induced droplet coalescence. Here, we explore numerically such type of complex merging schemes when one of the fluids is thixotropic.

\begin{figure*}[hbt!]
\centering
     \includegraphics[width=\textwidth]{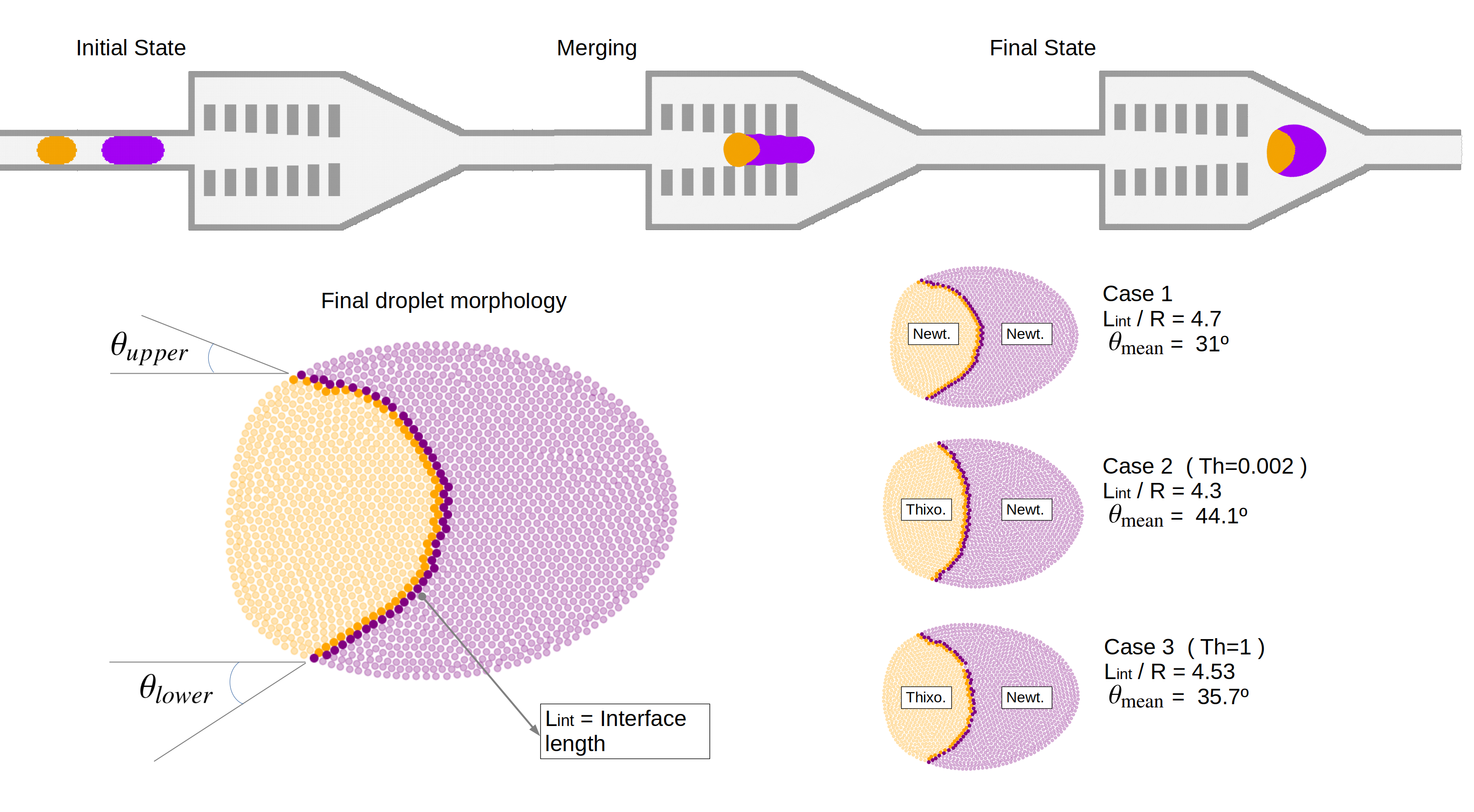}
\caption{Initial state and merging process using microfluidic device. For Case 1, both droplets were considered to be Newtonian. In Case 2 the recipient droplet (purple) is Newtonian and the penetrating droplet (orange) is thixotropic with $Th=0.002$. In Case 3 is the same that case 2 but $Th=1.0$.}
\label{fig:11}
\end{figure*}

In \Cref{fig:11}, we illustrate the merging domain configuration investigated. The physical boundaries (in gray) represent the channel walls and pillar structures. The continuous phase (in white) flows through the channel. We impose periodic boundary conditions in the  flow direction, and apply a body force to induce the flow.  We analyze three particular cases for which we set a sonic time scale $\tsnd= 0.14$, interfacial tension time scale $\tsig= 10$, viscous time scale $\tvis= 7.84$, characteristic flow time scale $\tconv= 490$ and thixotropic time $\tact=100$, following the hierarchy $\tsnd \ll \tvis, \tsig \ll \tconv, \tact$. In the first case, both the penetrating droplet (orange in \Cref{fig:11}) and the recipient droplet (purple) are Newtonian fluids with equal viscosity $\eta = 1$. In the second case, the penetrating droplet is thixotropic with properties $\eta_\infty = 1$, $\alpha = 9$ (this is $\eta_{\text{max}}=10\eta_\infty$), $\beta = 1$, and  $f_0=1$, while the recipient droplet remains Newtonian with $\eta = 1$. The third case has the same configuration as the second case, with the only difference being that $\beta=500$. In all cases, we consider the continuous phase as a Newtonian fluid with viscosity $\eta = 1$ and an input surface tension of $\sigma_0=2$. The height of the chamber in the central zone is equal to $H/\Delta x=220$ and the length $L/\Delta x=130$. The radius of the final droplet is calculated as $R/\Delta x = (\sqrt{N_{part}/d_{eq}}) / \pi = 20$. The maximum resultant velocity of the system is $U=0.09$ which results in a Capillary number of $Ca=0.045$, Reynolds number of $Re=2.4$ and a Thixotropic number of $Th=0.002$ and $Th=1$ for the two values of $\beta$.

In our simulation results, we observe that as the first droplet enters the chamber and experiences significant resistance due to its interaction with the pillar geometry. Meanwhile, the second droplet advances and begins to penetrate the first droplet, which is partially trapped within the pillar structure. During this interaction, the second droplet transfers part of its kinetic energy to the first, facilitating the joint passage of both droplets through the pillar region. Upon exiting the interaction zone, the smaller droplet is almost fully embedded within the larger one. This interaction results from surface tension, which promotes coalescence and stabilizes the new droplet structure formed after fusion. 

\Cref{fig:11} compares the geometrical features of the final merged droplets for the three cases analyzed. In particular, we analyze the variations in the interfacial length $L_{\text{int}}/R$ between droplets ($R$ is the initial radius of the droplets) and the mean contact angle, given by $\theta_{\text{mean}}=(\theta_{\text{upper}}+\theta_{\text{lower}})/2$. For the Newtonian-Newtonian merging (case 1), the penetrating droplet retains a semicircular cap shape inside the recipient, with an interfacial length $L_{\text{int}}/R = 4.7$, and mean angle $\theta_{\text{mean}} = 31^\circ$. In the thixotropic-Newtonian case with $\text{Th} = 0.002$ (case 2), the interfacial coverage is flatter, characterized by $L_{\text{int}}/R = 4.3$ and $\theta_{\text{mean}} = 44.1^\circ$. In the case 3, with $\text{Th}=1$, the interface adopts an intermediate configuration with $L_{\text{int}}/R = 4.53$ and $\theta_{\text{mean}} = 35.7^\circ$.

The differences between the three cases reflect how the hierarchy of time scales controls the dynamics of fusion. In all systems, $\tsig \ll \tact \ll \tconv$ holds, so that interface formation occurs much faster than microstructural reconstruction, which in turn occurs faster than global convective transport. In the case of low Th, the evolution of the microstructure is slow, retaining a higher viscosity during the merging and resulting in a flatter interface. Conversely, when Th increases, microstructural destruction occurs on a much smaller time scale, reducing the effective viscosity before contact and favoring deeper and more symmetrical penetration, similar to that observed in the Newtonian case. Taken together, these results show that the relationship between the scales $\tact$ and $\tconv$ determines the dominant regime of the process. When $\tact \ll \tconv$, the drop tends to behave like a more viscous fluid that is resistant to deformation, while, as $\tact$ increases or internal relaxation accelerates (higher Th), coalescence becomes faster and the interface adopts a greater curvature. These findings highlight two potential implications of thixotropic fluids in droplet-merging related process. First, the breadth and form of droplet fusion can be controlled by tuning factors the microstructural evolution of the fluid, offering an adjustable delay in the merging dynamics. Second, the difference between flat to circular penetration patterns illustrates the dynamic variation of the effective contact angle, which can be used in microfluidic applications that require controlled release, encapsulation, or mixing. 

\section{Conclusions}
\label{con}

We proposed a numerical methodology based on Smoothed Dissipative Particle Dynamics (SDPD) for simulating multiphase flows with thixotropic behavior. The model accounts for surface tension and tracks the time-dependent evolution of viscosity due to microstructural changes. The approach was validated through static and dynamic cases, including droplet deformation, contact angles, Poiseuille flow, and flow past obstacles. It accurately reproduced known behaviors and matched analytical predictions, confirming its reliability.

The validated thixotropic multiphase SDPD model was applied to a range of complex flow scenarios, demonstrating its potential for investigating biologically and industrially relevant systems. In the case of liquid–liquid phase separation (LLPS), the model captured how parameters such as protein concentration, the capillary-driven coalescence, and microstructural relaxation influence droplet morphology—revealing conditions that favor the formation of stable, well-defined aggregates. Simulations of suspensions and emulsions showed that the model accurately reproduces the rheological evolution of complex fluids under confinement and shear, including effective viscosity trends and interfacial dynamics in two-phase systems.
\\\\
In microfluidic geometries, such as periodically constricted channels and droplet merging devices, the model captured key behaviors like localized viscosity variation, flow stabilization, and coalescence resistance due to internal structure. Moreover, this results illustrate how tuning both the rheological parameters and the geometric design, it becomes possible to tailor viscosity profiles, shear stresses, and relaxation dynamics to specific functional needs. This coupling provides a framework for designing devices that exploit thixotropy, such as microfluidic systems for drug delivery, flow regulators in biomedical contexts, or extrusion processes in manufacturing.
\\\\
Overall, our numerical results highlight the model’s capacity to simulate time-dependent rheology and interface behavior, making it a promising tool for the design and analysis of systems in soft matter physics, biomedical applications, and microfluidic device engineering.

\section*{Acknowledgments}
The authors acknowledge funding by the Basque Government through the BERC 2022-2025 program and by the Ministry of Science, Innovation and Universities: BCAM Severo Ochoa accreditation CEX2021-001142-S / MICIN / AEI /10.13039/ 501100011033. The Spanish State Research Agency through the project PID2020-117080RB-C55 MICIU/AEI /10.13039 / 501100011033 funded by (AEI/FEDER, UE) with acronym COMPU -~NANO~-~HYDRO. 

\section*{CRediT authorship contribution statement}
\textbf{Andres Santiago Espinosa-Moreno:} Investigation, Software, Validation, Writing – original draft, Writing – review \& editing, Formal analysis. \textbf{Nicolas Moreno:} Conceptualization, Investigation, Supervision, Software, Methodology, Writing – original draft, Writing – review \& editing, Formal analysis. \textbf{Marco Ellero:} Writing – review \& editing, Supervision, Resources, Methodology, Funding acquisition, Formal analysis, Conceptualization.

\newpage
\appendix
\section{Simple Newtonian Flows}
\label{app1}

\begin{figure*}[htbp!]
\centering
     \begin{subfigure}[b]{0.32\textwidth}
         \centering
     \includegraphics[width=\textwidth]{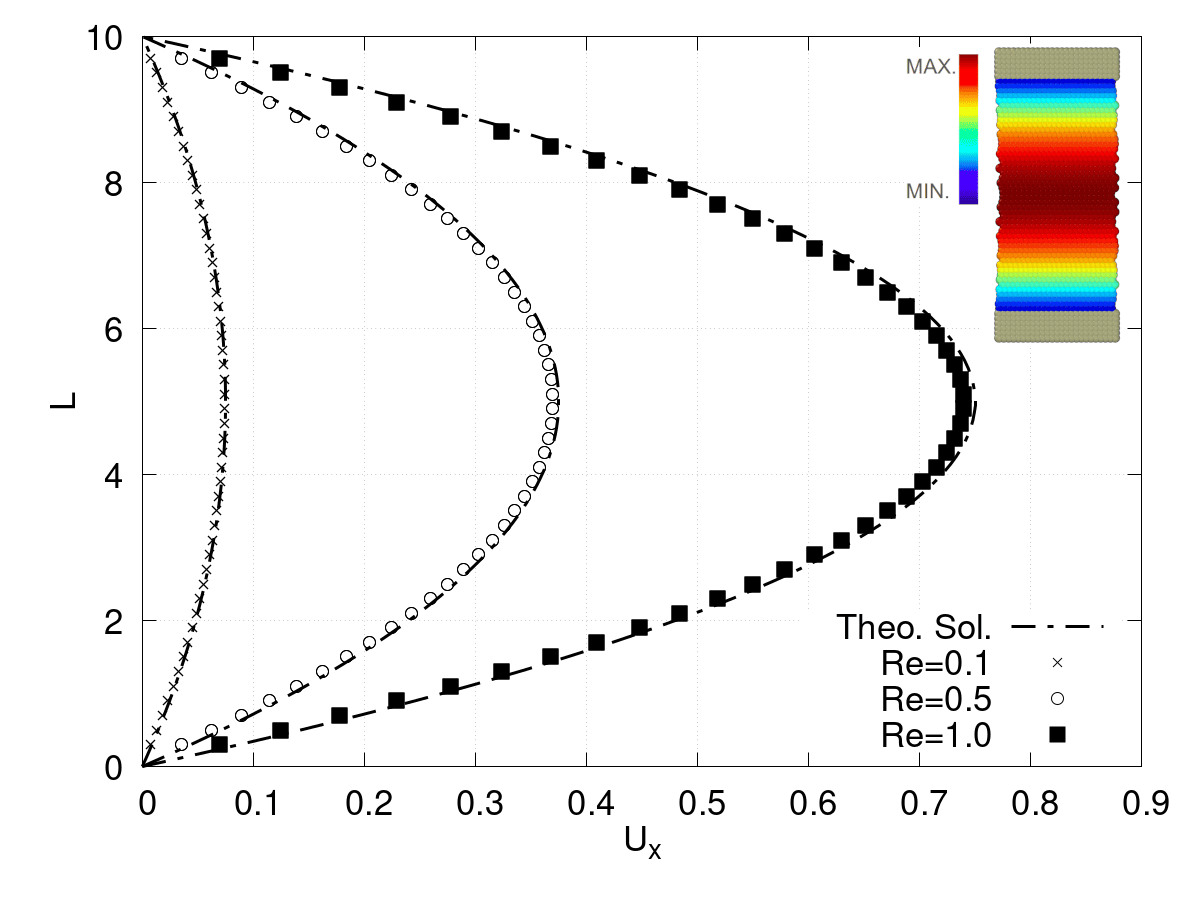}
         \caption{Simple Poiseuille velocity profiles}
         \label{fig:A12-a}
     \end{subfigure}
     \begin{subfigure}[b]{0.32\textwidth}
         \centering
     \includegraphics[width=\textwidth]{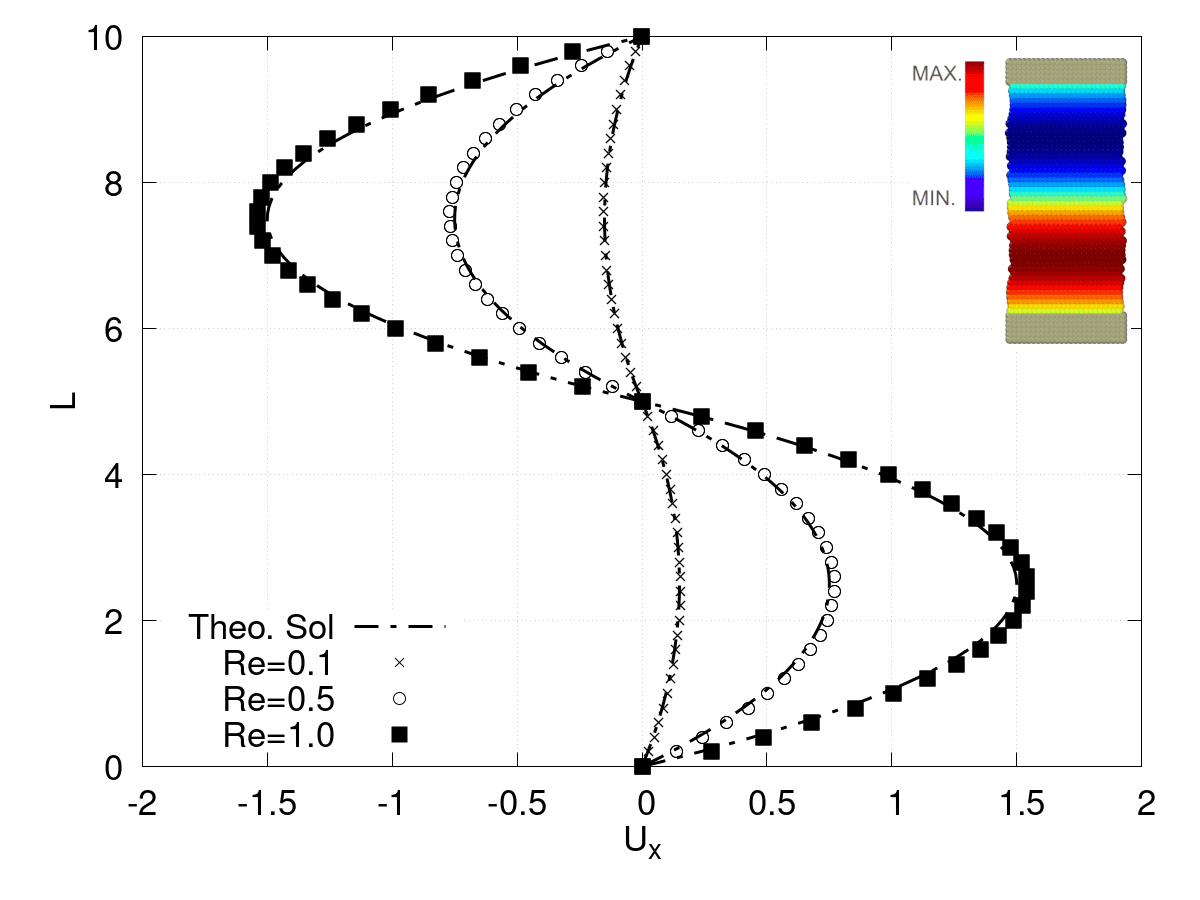}
         \caption{Reverse Poiseuille velocity profiles}
         \label{fig:A12-b}
     \end{subfigure}
     \begin{subfigure}[b]{0.32\textwidth}
         \centering
     \includegraphics[width=\textwidth]{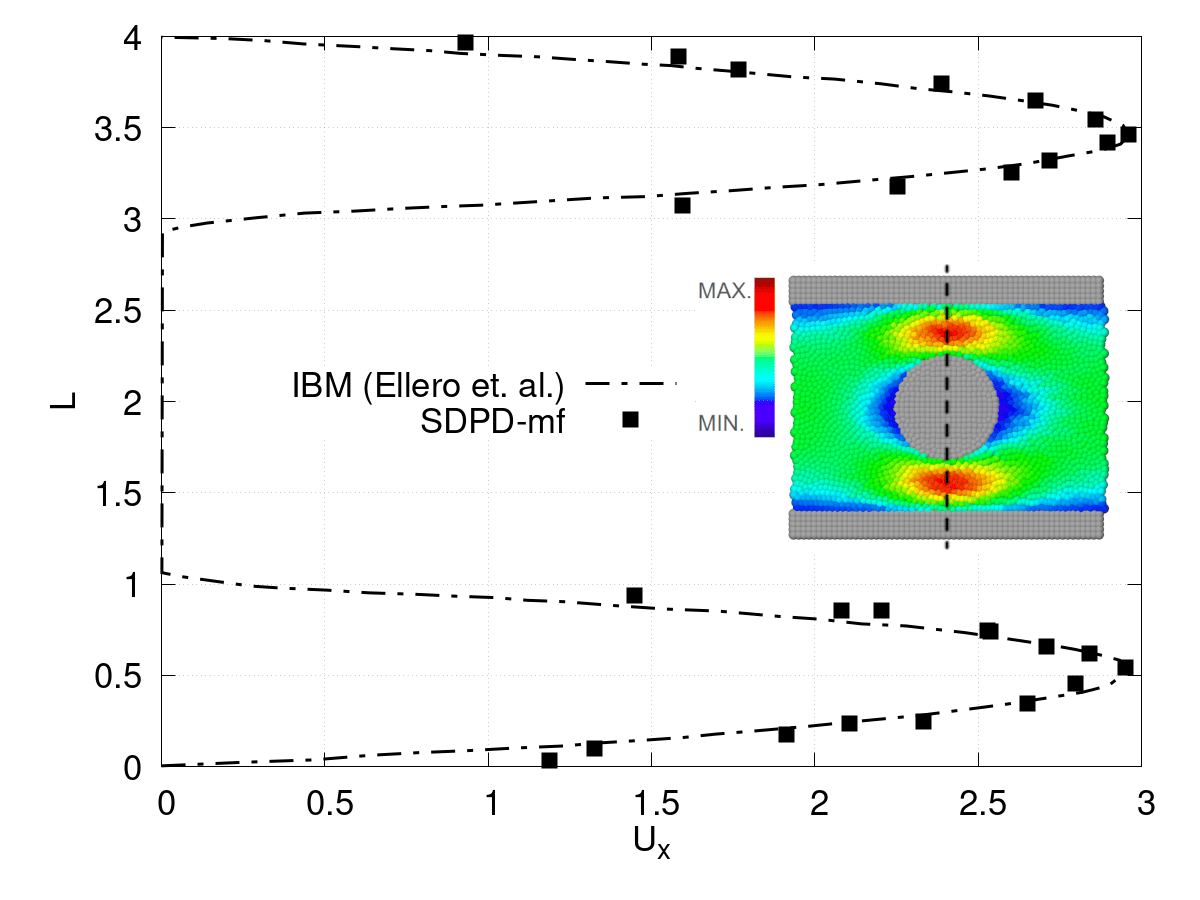}
         \caption{Flow around a cylinder. Section A.}
         \label{fig:A12-c}
     \end{subfigure}

     \begin{subfigure}[b]{0.32\textwidth}
         \centering
     \includegraphics[width=\textwidth]{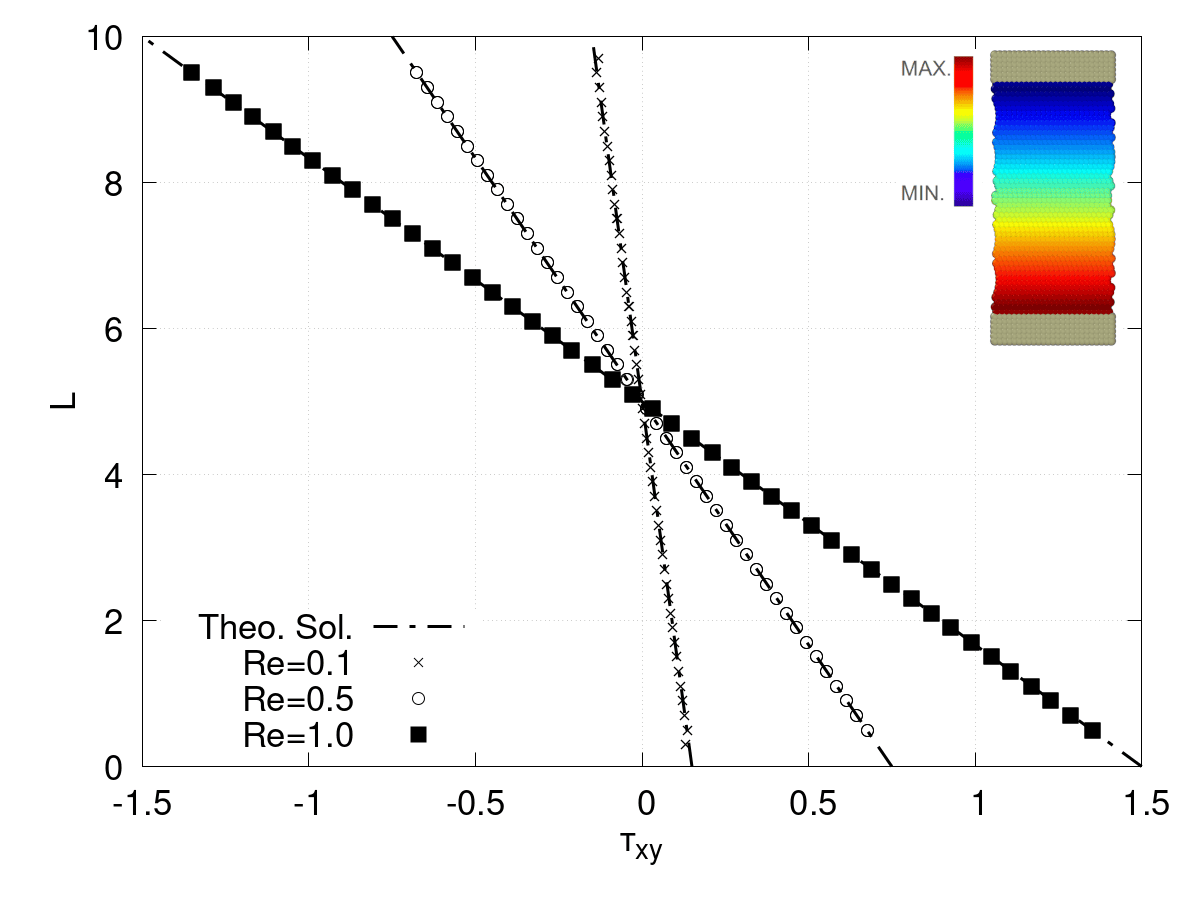}
         \caption{Simple Poiseuille shear stress profiles}
         \label{fig:A12-d}
     \end{subfigure}
     \begin{subfigure}[b]{0.32\textwidth}
         \centering
     \includegraphics[width=\textwidth]{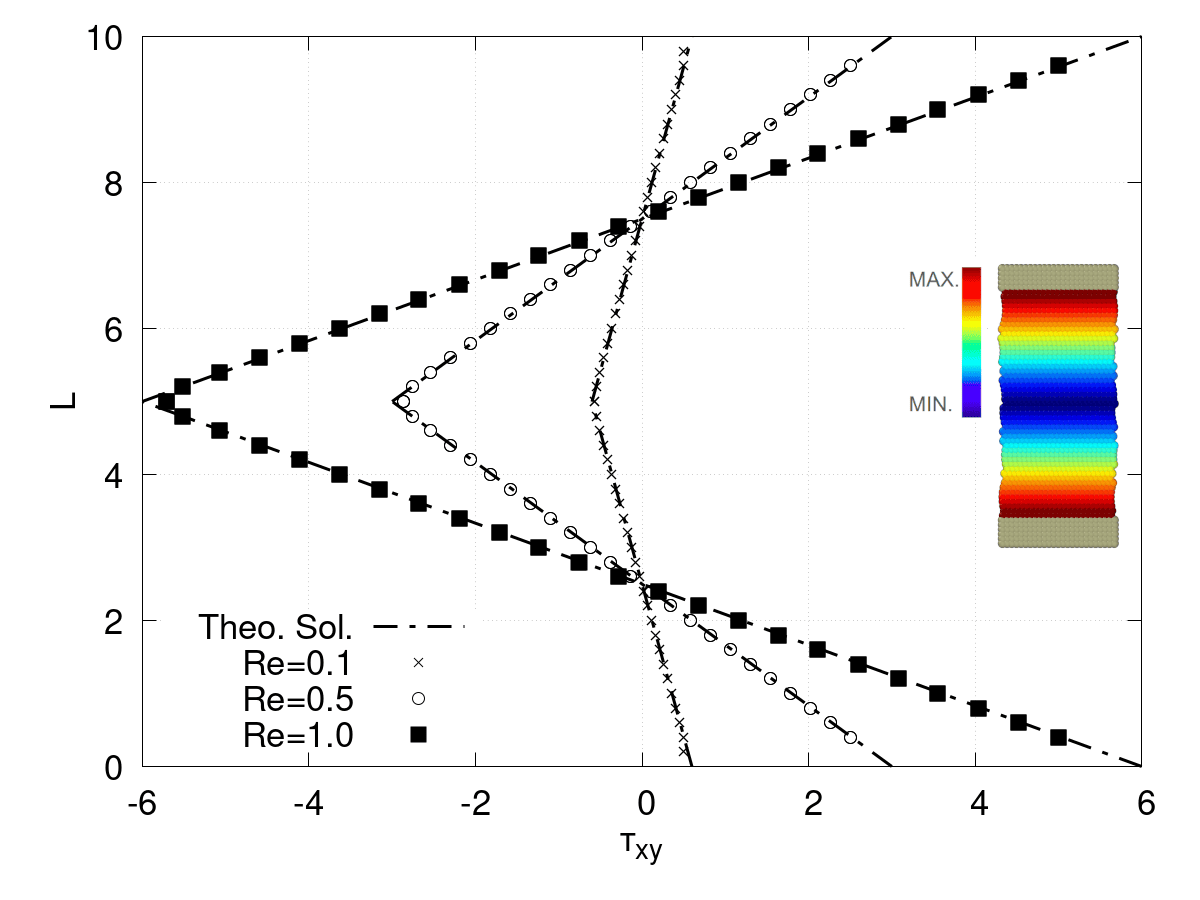}
         \caption{Reverse Poiseuille  shear stress profiles}
         \label{fig:A12-e}
     \end{subfigure}
     \begin{subfigure}[b]{0.32\textwidth}
         \centering
     \includegraphics[width=\textwidth]{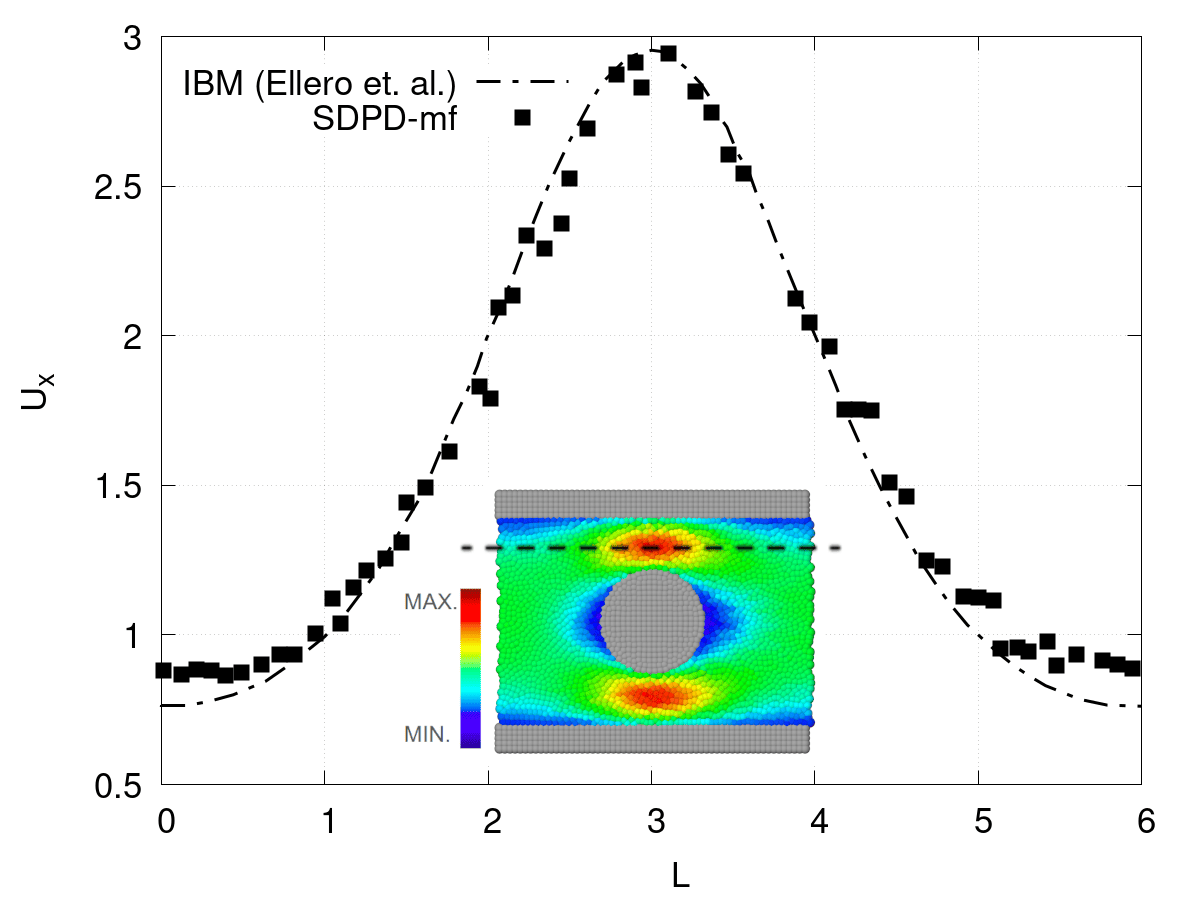}
         \caption{Flow around a cylinder. Section B}
         \label{fig:A12-f}
     \end{subfigure}
     
\caption{Validation of the methodology for one phase  Poiseuille flows using (a)-(d) simple profile, (b)-(e) reverse profile and (c)-(f) flow around a cylinder in a vertical and horizontal plane. }
\label{fig:A12}
\end{figure*}

We carried out the dynamic validation of the model considering several scenarios. The first part of the validation involves the development of a Poiseuille flow for a single-phase flow ($\alpha$) in three variations: simple profile, reverse Poiseuille and flow around a cylinder. For particle methods, as indicated by Backer et al. \cite{backer2005poiseuille}, the average velocity and shear stress in a Poiseuille flow are given by
\begin{align}\label{eq:pou}
    v_{ave} &= <v_x> = \frac{1}{L} \int_0^L v_x(y) dy = \frac{\rho g_x L^2}{12 \eta}, \\
    \tau_{xy} &= \rho g_x \left( y - \frac{1}{2} L \right),
\end{align}
where $\rho$ is the density, $L$ the length between the parallel walls and $g_x$ the body force. For the simple and reverse case, we employ a two-dimensional square channel with walls separated by $L/\Delta x=50$, $\rho=1$ and $\eta=5$ in the units system. We select the equilibrium particle density $n_{eq}=25$ and we do not taken into account the effects of thermal fluctuations. We compare three velocities under laminar flow regime determine by the Reynolds number $Re=\rho v_x D / \eta$. We configure the body force with the required values to obtain the respective predefined velocities. In the case of reverse Poiseuille flow, the body force is positive in one half of the channel and negative in the opposite half, hence resulting in the associated representative velocity profile. We choose the speed of sound to be at least 20 times greater than the maximum velocity reached by the flow. We established a periodicity condition along the x-direction and we define a no-slip condition for walls by setting the related particles as totally stationary and without any contribution to the fluid. \Cref{fig:A12-a,fig:A12-b,fig:A12-d,fig:A12-e} shows the velocity and shear profiles calculate by our model in comparison to the theoretical values obtain by applying \Cref{eq:pou} for $Re=[0.1,0.5,1]$. Regarding the case of flow around a cylinder, we perform a validation based on the numerical results obtained by Ellero and Adams \cite{ellero2011sph} using both SPH and immersed boundary method (IBM). We employ a channel characterized by a centrally located cylinder with radius $R_c/\Delta x = 10$ and height $L=4 R_c$ to study the flow characteristics. Considering the condition of periodicity on the x-direction, we set the width of the channel $L_c=6 R_c$ such that the distance between repeated cylinders allows the fluid development. For this validation, the same physical parameters used in the first two cases are kept with the only exception of the viscosity whose new value is $\eta=1$ (in line with the approach of Ellero and Adams \cite{ellero2011sph}). We configure the cylinder as well as the channel walls with a no-slip condition. \Cref{fig:A12-c,fig:A12-f} shows the velocity profiles versus the normalized $x/R_c$ and $y/R_c$ axis for both for the results obtained with our approach and those getting by the IBM \cite{ellero2011sph}. These curves are taken on a vertical plane at a distance of $x/R_c=3$ and a horizontal plane at a distance of $y/R_c=3.5$, respectively. As can be noted for the three analyses cases of single-phase flow, the comparative results between the theoretical values and the numerical approximations exhibit a close agreement showing the accuracy of the model in order to capture the characteristics of the flow under dynamic conditions.

\section{Droplet size distribution}
\label{app2}

\begin{figure*}[htbp!]
    \centering
         \begin{subfigure}[b]{0.45\textwidth}
             \centering
         \includegraphics[width=\textwidth]{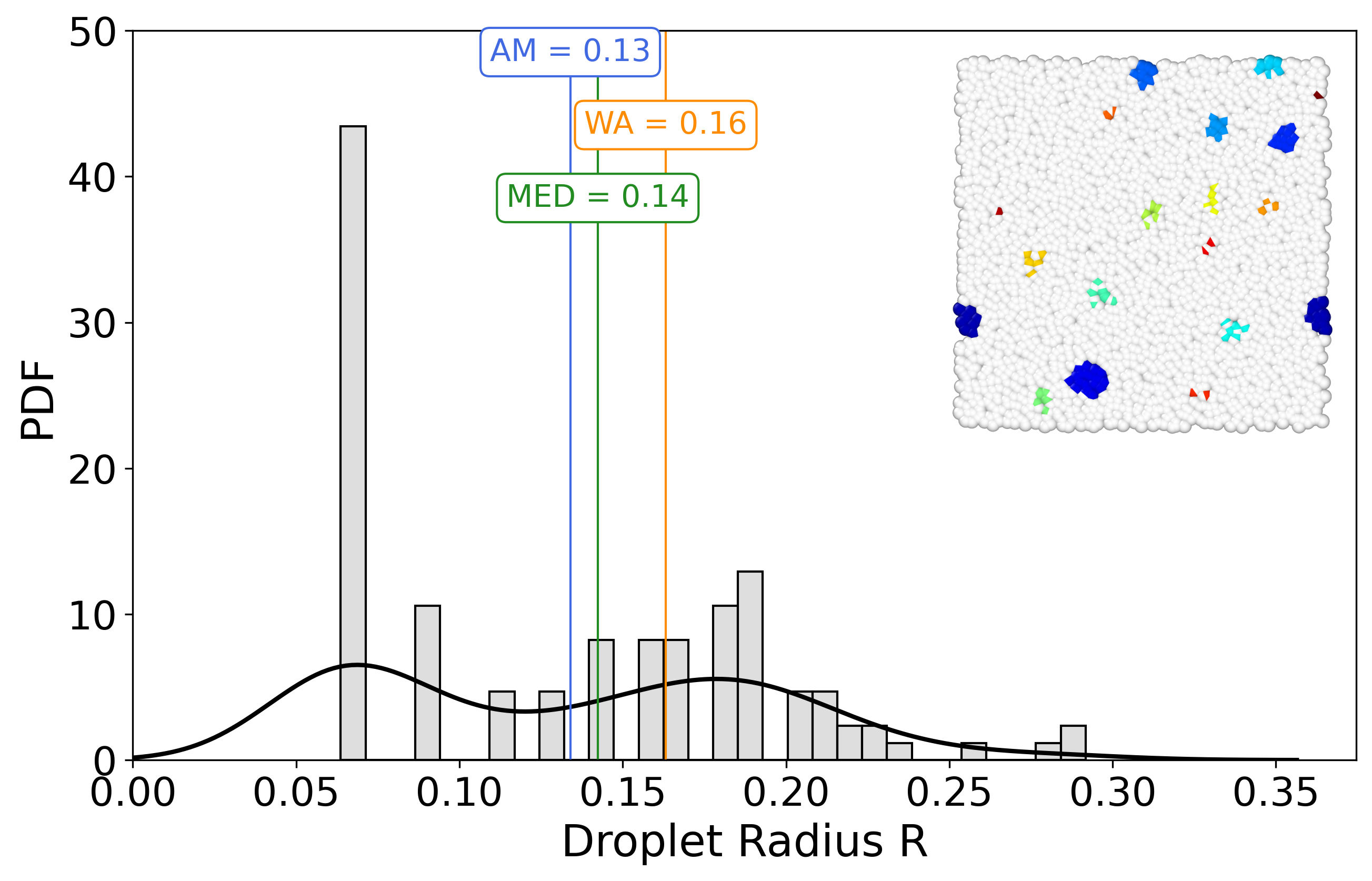}
             \caption{Protein phase concentration $\Phi_d=5\%$}
             \label{fig:B14-a}
         \end{subfigure}
         \begin{subfigure}[b]{0.45\textwidth}
             \centering
         \includegraphics[width=\textwidth]{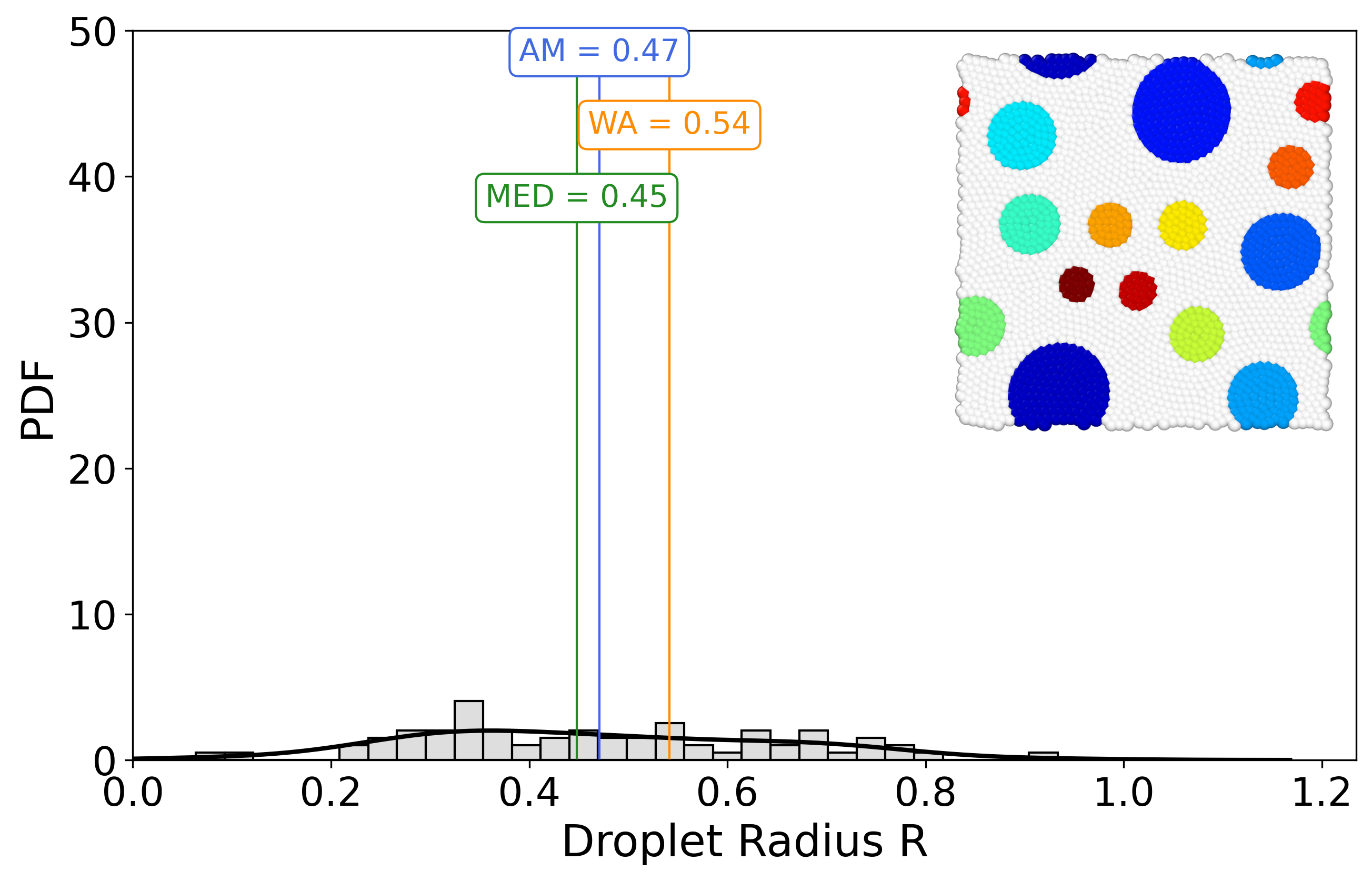}
             \caption{Protein phase concentration $\Phi_d=35\%$}
             \label{fig:B14-b}
         \end{subfigure}
         \begin{subfigure}[b]{0.45\textwidth}
             \centering
         \includegraphics[width=\textwidth]{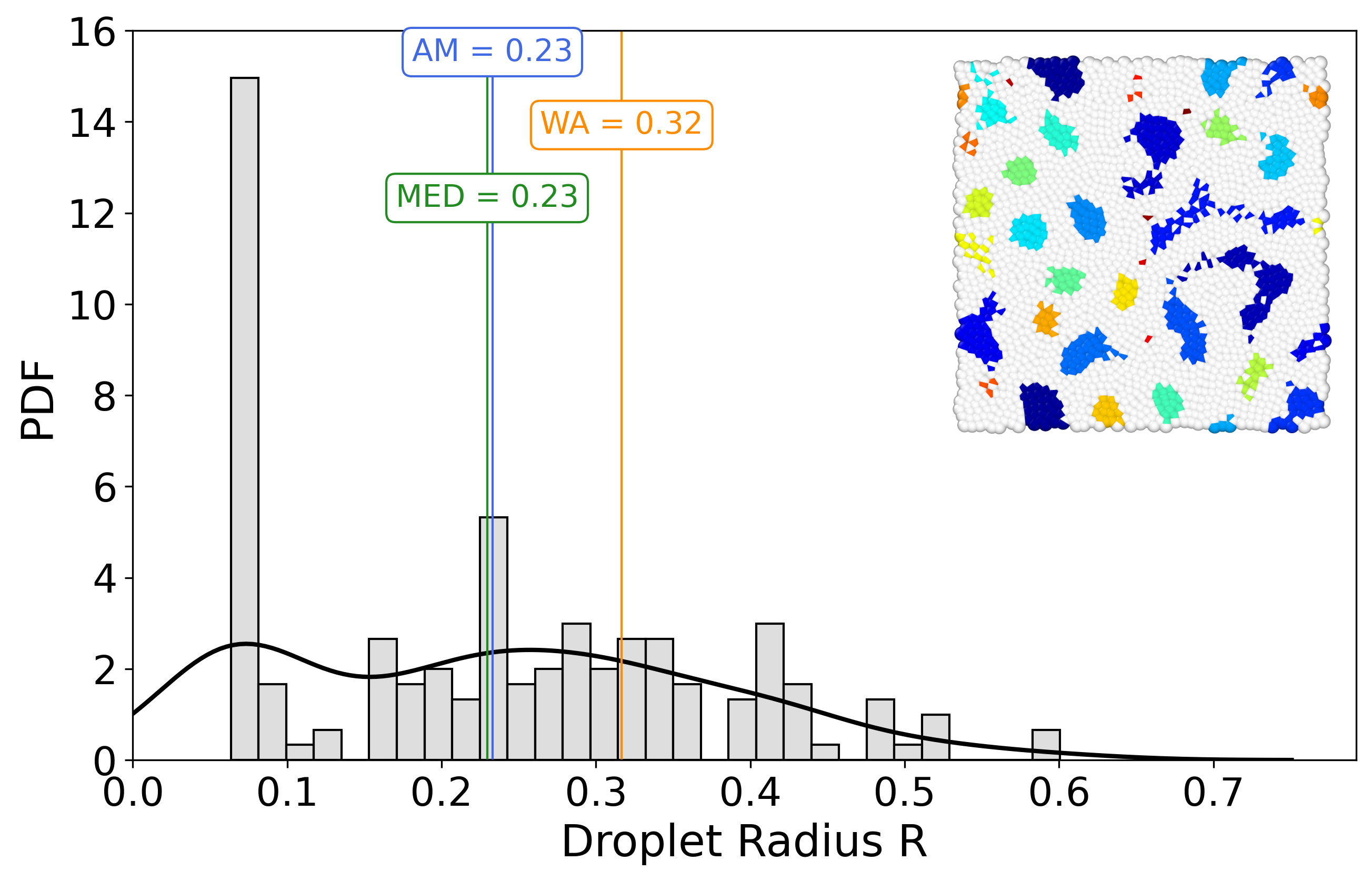}
             \caption{Ratio $\tact/\tsig=0.1$}
             \label{fig:B14-c}
         \end{subfigure}
         \begin{subfigure}[b]{0.45\textwidth}
             \centering
         \includegraphics[width=\textwidth]{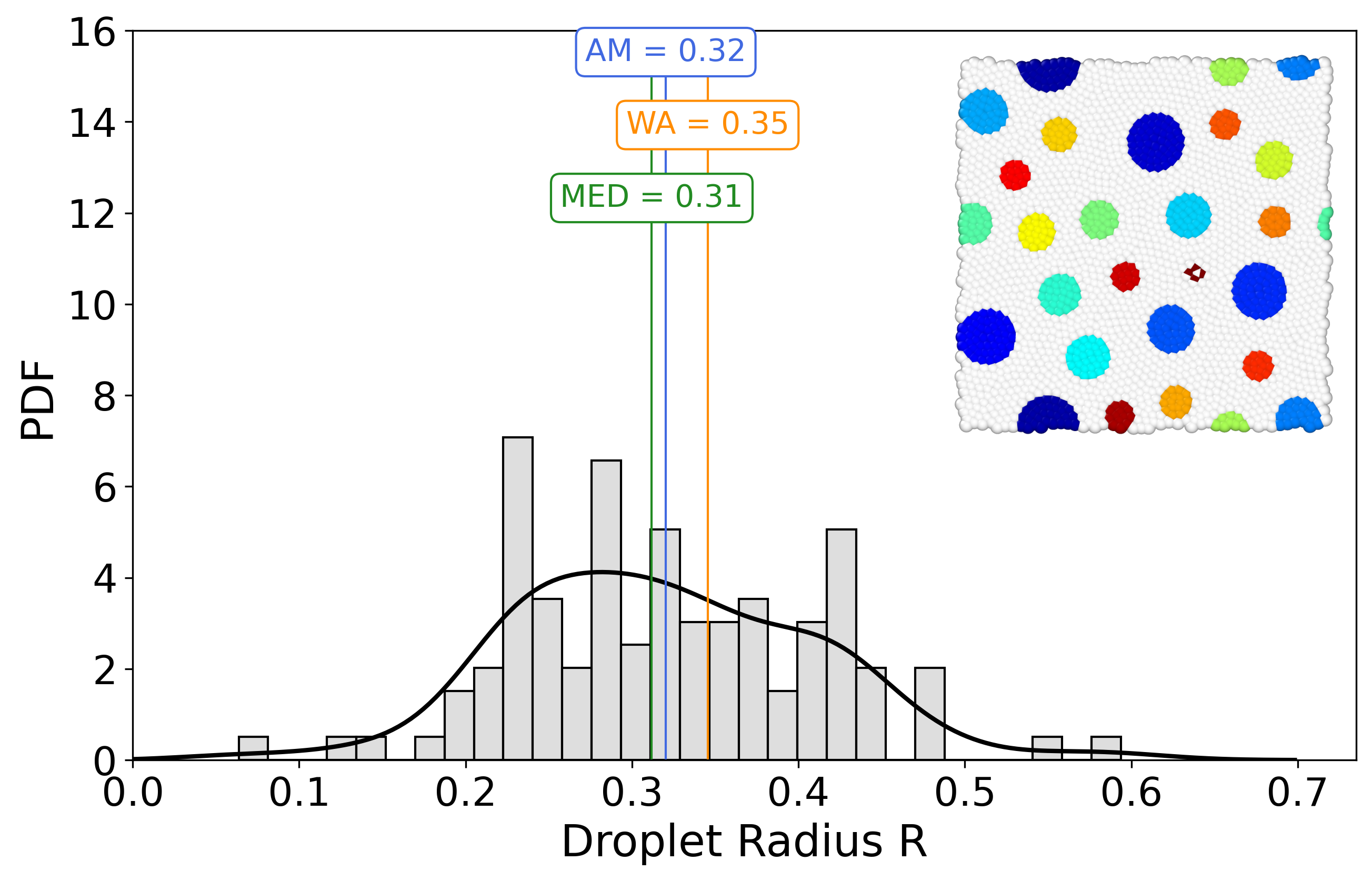}
             \caption{Ratio $\tact/\tsig=2$}
             \label{fig:B14-d}
         \end{subfigure}
              \begin{subfigure}[b]{0.45\textwidth}
             \centering
         \includegraphics[width=\textwidth]{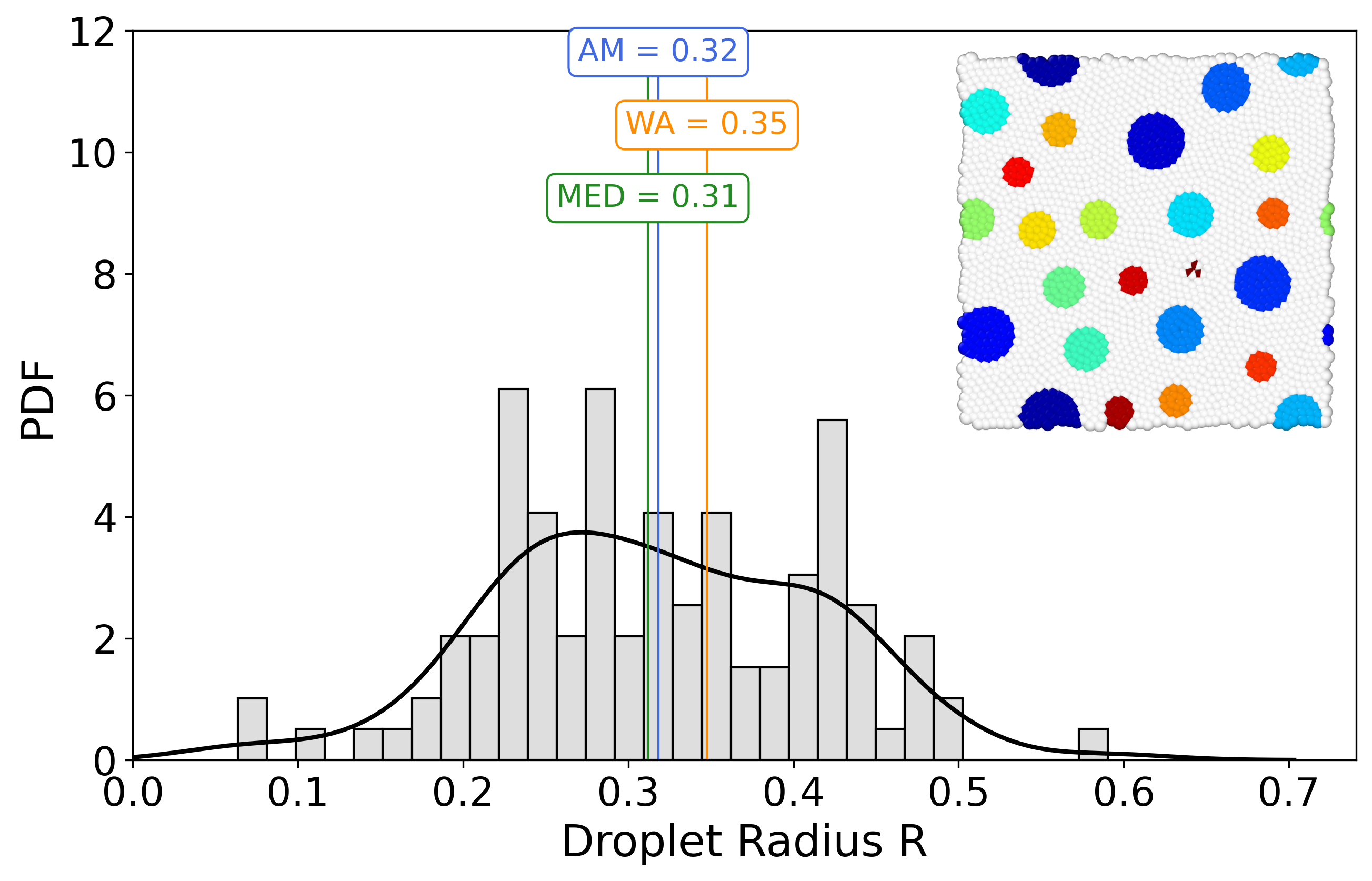}
             \caption{Constitutive constant $\frac{1}{1+\alpha} = 0.5$}
             \label{fig:B14-e}
         \end{subfigure}
         \begin{subfigure}[b]{0.45\textwidth}
             \centering
         \includegraphics[width=\textwidth]{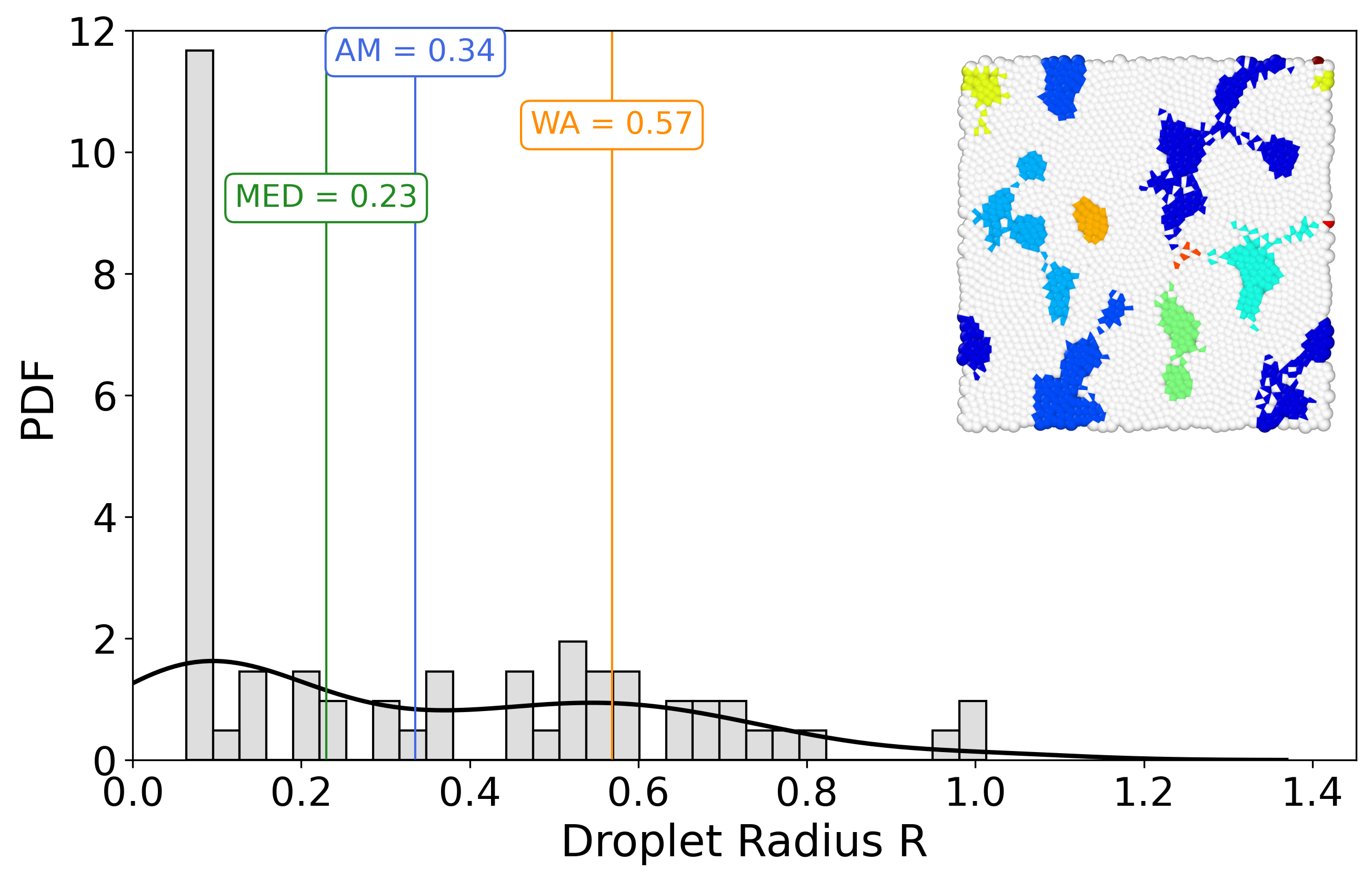}
             \caption{Constitutive constant $\frac{1}{1+\alpha} = 0.01$}
             \label{fig:B14-f}
         \end{subfigure}
              \begin{subfigure}[b]{0.45\textwidth}
             \centering
         \includegraphics[width=\textwidth]{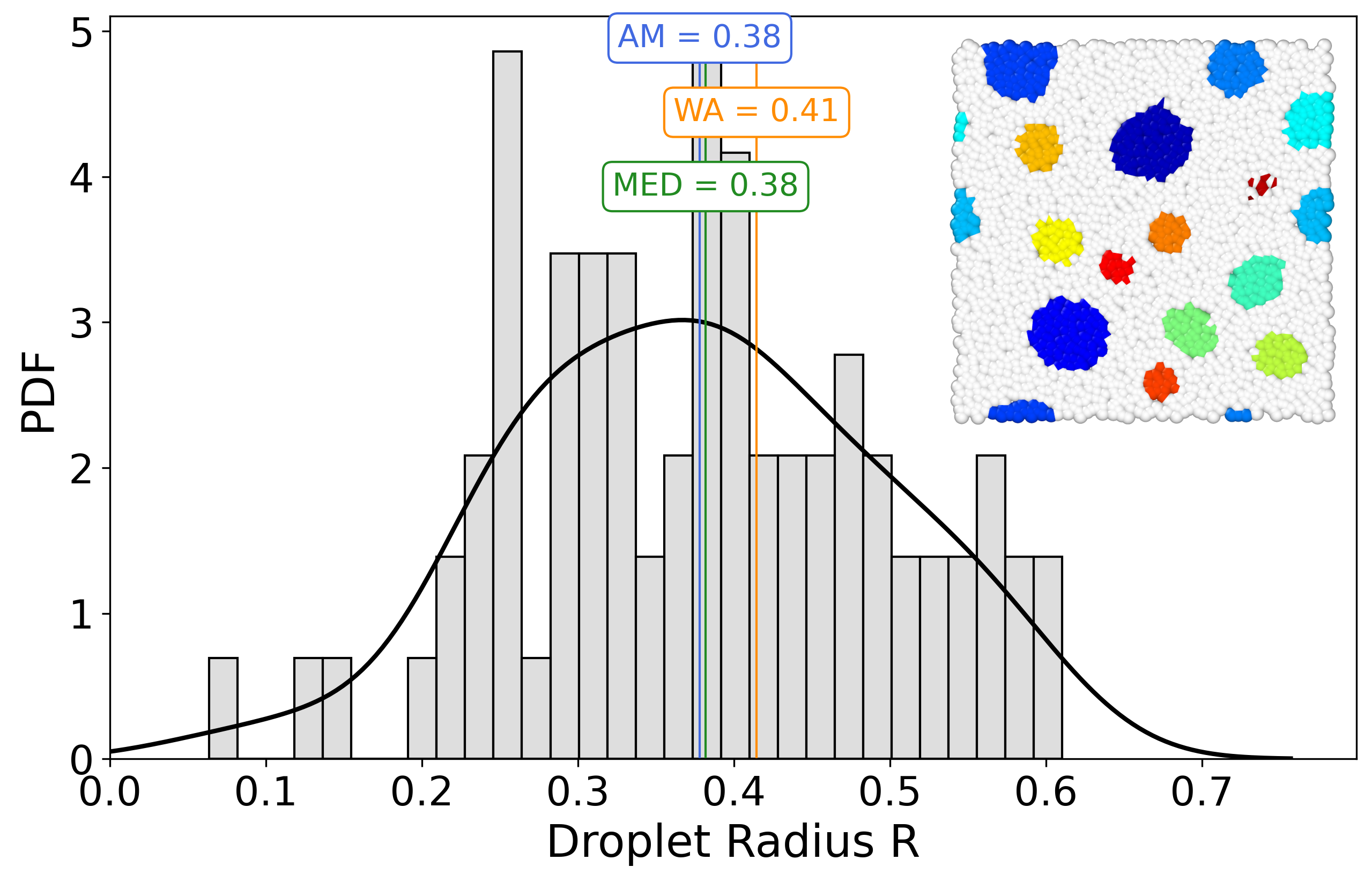}
             \caption{Thermal fluctuations $k_BT = 0.05$}
             \label{fig:B14-g}
         \end{subfigure}
         \begin{subfigure}[b]{0.45\textwidth}
             \centering
         \includegraphics[width=\textwidth]{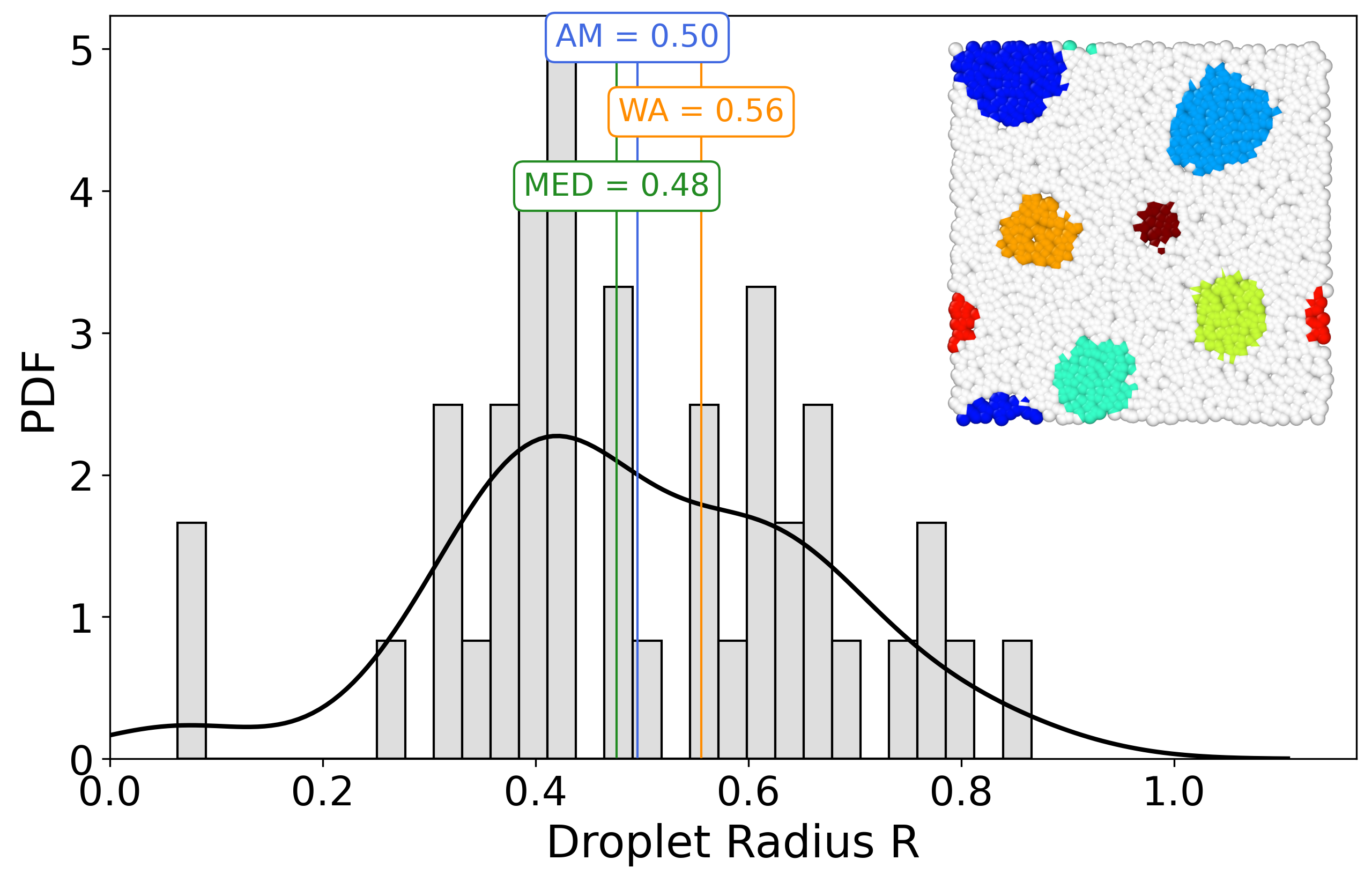}
             \caption{Thermal fluctuations $k_BT = 0.1$}
             \label{fig:B14-h}
         \end{subfigure} 
         
    \caption{Cluster histograms for the comparison of (a) protein phase concentration, (b) variation of the ratio between the thixotropic characteristic time and interfacial tension time scale $\tact/\tsig$, (c) constitutive constant ${1}/{(1+\alpha)}$, and (d) thermal energy $k_B T$.}
    \label{fig:B14}
    \end{figure*}

We perform numerical simulations using the multiphase SDPD methodology proposed in this work for the analysis of LLPS phenomena and examined four main properties: (i) protein phase volume fraction $\Phi_d$, (ii) variation of the ratio between the thixotropic characteristic time and interfacial tension time scale $\tact/\tsig$, (iii) constitutive constant ${1}/{(1+\alpha)}$, and (iv) thermal energy $k_B T$. For each property we construct histograms and estimate the probability density using the KDE function. In addition, we compute the Probability Density Function (PDF) of the droplets size, and introduce a set of relevant metrics, such as the average molecular weight (AM), the median (MED), the weighted-average molecular weight (WA), and polydispersity index (PDI). These metrics are defined as follows
%
\begin{equation*}
AM =  \frac{\sum_i^{N_\text{drop}} N_iM_i}{\sum_i N_i}, \quad WA =  \frac{\sum_i^{N_\text{drop}} N_iM_i^2}{\sum_i N_iM_i}, \quad PDI = \frac{Mw}{Mn},
\end{equation*}
where $N_{\text{drop}}  = \sum N_i$ and $M_i$ is the size of the $i$-th droplet. In this appendix we present the histograms corresponding to each property, accompanied by a representative snapshot of the last simulated time. The results for protein phase volume fraction $\Phi_d=[5\%,35\%]$,  $\tact/\tsig =\{0.1, 0.5, 1, 2\}$, ${1}/{(1+\alpha)}=[0.5,0.01]$ and $k_B T=[0.05,0.1]$ are shown in \Cref{fig:B14}.

\begin{figure}[htbp!]
    \centering
     \includegraphics[width=0.47\textwidth]{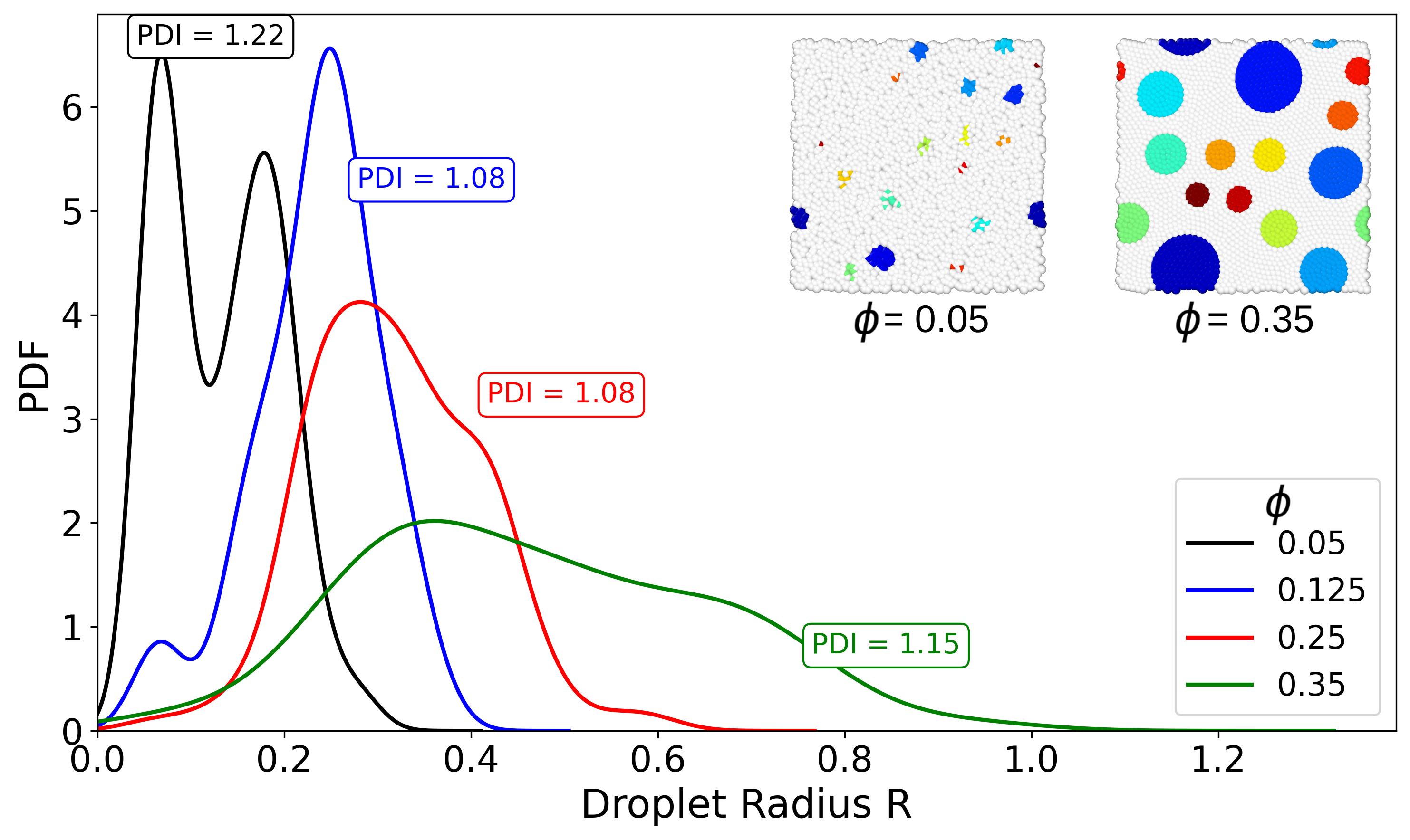}
    \caption{KDE comparison for four different protein concentration including the PID value.  Inside snapshot illustrate the final configuration for two different system, where droplets are depicted in different colors for clarity.}
    \label{fig:B13}
\end{figure}

In \Cref{fig:B13}, we compile the different Kernel Density Estimation (KDE) curve obtained for the different volume fractions. An snapshot of the final state for two concentration is included. In \ref{app2}  (\Cref{fig:B14-a} and \Cref{fig:B14-b}), we provide an example of droplet sizes histograms for $\Phi_d=5\%$ and $\Phi_d=35\%$, illustrating the central tendency statistics. The results indicate that at the lowest volume fractions ($\Phi_d=5\%$) investigated, interfacial tension effects are not sufficient to drive the formation of well-defined droplets. In contrast, at higher volume fractions ($\Phi_d=35\%$), the protein phase exhibits a tendency to form regular circular droplets, driven by interfacial tension phenomena.  Overall, at lower proteins content lead to greater heterogeneity (broader droplet size), whereas as the volume increases, the droplet size distribution tend exhibit lower dispersions.

\section{Effective viscosity in suspensions} \label{app3}

\begin{figure}[hbt!]
\centering
     \includegraphics[width=0.5\textwidth]{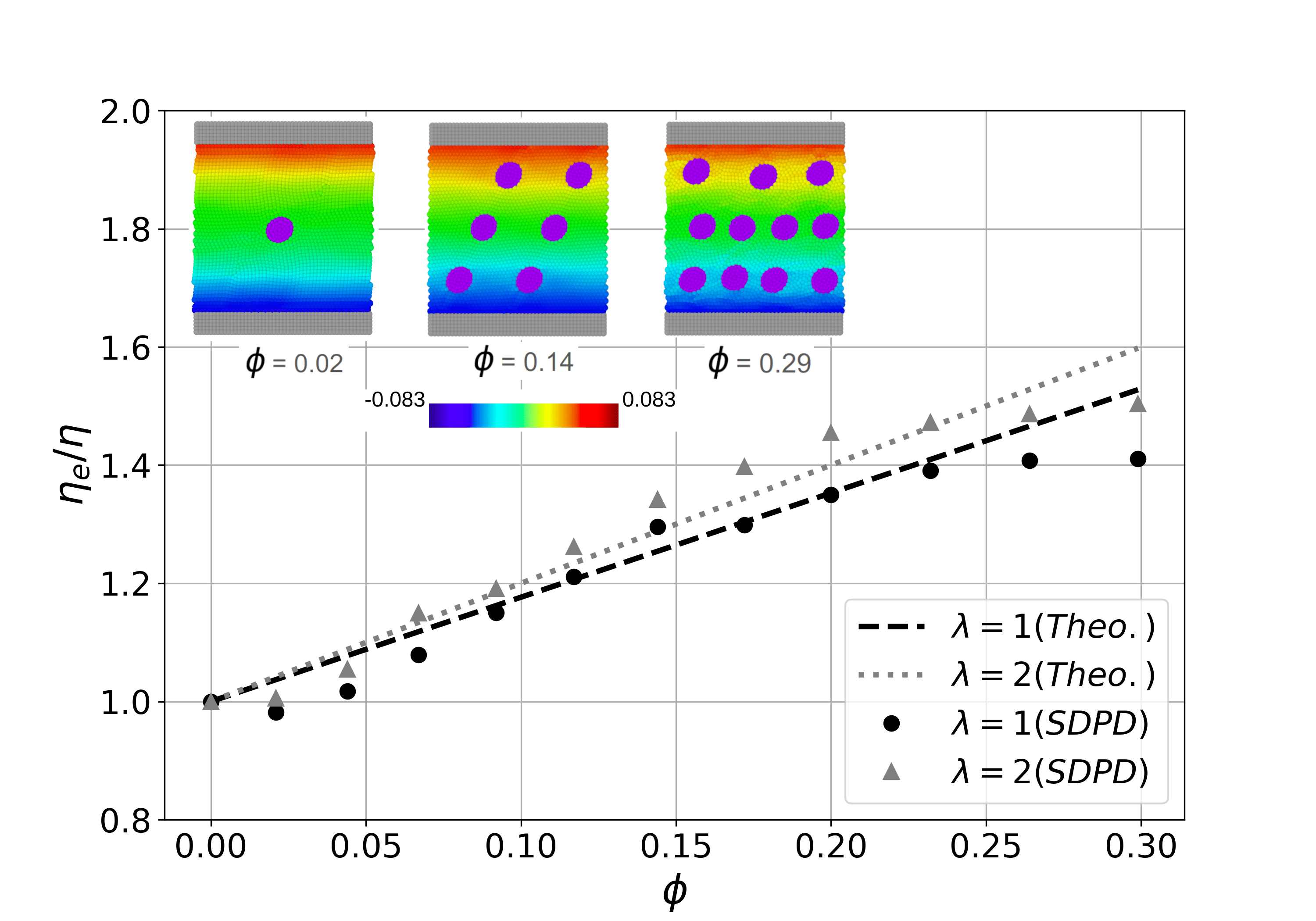} 
\caption{Emulsion validation using Taylor`s expression for two viscosity ratios $\lambda$}
\label{fig:C-15}
\end{figure}

The behavior of rigid particles suspended in a fluid was studied by Einstein \cite{einstein1906}, where the expression $\eta_{eff} = \eta (1 + B\phi)$ was derived. This indicates that for concentrations less than 1 ($\phi <<1$) the effective viscosity of the particle suspension is directly proportional to the viscosity of the fluid and the volume fraction of rigid particles, with a constant value $B = 2.5$. For deformable particles, Taylor \cite{taylor1932viscosity} proposed:

\begin{equation}\label{neff}
\eta_{eff} = \eta \left[ 1 + \left( \frac{D+2}{2} \right) \phi \left( \frac{\eta' + 0.4 \eta}{\eta' + \eta} \right) \right]
\end{equation}

where $\eta'$ is the viscosity of the dispersed phase and $D$ is the system dimension. When $\eta' = \eta$, this simplifies to $\eta_{eff} = \eta (1 + 1.75\phi)$. This model assumes moderate droplet deformability, i.e., low Capillary numbers. 
We conduct simulations in a square channel of size $L/\Delta x = 50$ under simple shear flow $\dot{\gamma} = 0.0166$, and $k_BT\to 0$. The continuous phase is Newtonian with $\rho = 1$, $\eta = 15$ and surface tension $\sigma_0 = 2$. Droplets have radius $r_0 = 0.8$ and viscosities $\eta' = \{15 \text{ and } 30\}$ covering two viscosity ratios of $\lambda = \{1 \text{ and } 2\}$. This leads to $Re = 1.5 \times 10^{-3}$ and $Ca = 0.1$.  We perform a total of eleven simulation with volume fractions in the range $0 < \phi < 0.3$ by adjusting the number of droplets. The effective viscosity is calculated by direct measurement of the force on the wall. \Cref{fig:C-15} shows the results, including snapshots for $\phi = [0.02, 0.14, 0.29]$, illustrating droplets deformation. The model captures the increasing trend in viscosity, followed by a slight drop for volume fraction exceeding 0.2, consistent with prior observations \cite{caserta2012vorticity,de2019effect,rosti2021shear,rosti2019numerical}.

\end{document}